\documentclass{article}

\usepackage[a4paper, total={6.1in, 8.1in}]{geometry}
\usepackage[round]{natbib}
\usepackage{authblk}
\usepackage{amssymb}
\usepackage{amsmath}
\usepackage[utf8]{inputenc}
\usepackage[T1]{fontenc}
\usepackage{xcolor}
\definecolor{linkcolor}{rgb}{0,0,0.5} 

\usepackage[colorlinks=true,
    urlcolor=linkcolor, 
	linkcolor= linkcolor, 
	citecolor=linkcolor 
]{hyperref}
\usepackage{graphicx}

\usepackage{subfig}

\usepackage{cleveref}
\crefname{section}{\S}{\S}
\crefname{subsection}{\S}{\S}
\crefname{subsubsection}{\S}{\S}

\graphicspath{ 
	      {FiguresHL/}}

\newtheorem{rmk}{Remark}

\newcommand{\tI}[1]{\mathbf{#1}}
\newcommand{\tII}[1]{\mathcal{\mathbf{#1}}	}
\newcommand{\ie}{\emph{i.e.}~}
\newcommand{\eg}{\emph{e.g.}~}

\title{Development and validation of a numerical wave tank based on the Harmonic Polynomial Cell and Immersed Boundary methods to model nonlinear wave-structure interaction}

\author[1]{Fabien Robaux}
\author[1,2]{Michel Benoit} 

\affil[1]{\href{mailto:robaux@irphe.univ-mrs.fr}{robaux@irphe.univ-mrs.fr}, Aix Marseille Univ, CNRS, Centrale Marseille, Institut de Recherche sur les Ph\'enom\`enes Hors-Equilibre (IRPHE), Marseille, France}
\affil[2]{\href{mailto:benoit@irphe.univ-mrs.fr}{benoit@irphe.univ-mrs.fr}, Centrale Marseille, Marseille, France}

\date{\today}

\begin{document}

\maketitle
Accepted manuscript for publication, \citet{robaux_development_2021-1}, \emph{Journal of Computational Physics}, ISSN 0021-9991.

\begin{abstract}
    A fully nonlinear potential Numerical Wave Tank (NWT) is developed in two dimensions, using a combination of the Harmonic Polynomial Cell (HPC) method for solving the Laplace problem on the wave potential and the Immersed Boundary Method (IBM) for capturing the free surface motion. This NWT can consider fixed, submerged or wall-sided surface piercing, bodies. 
    To compute the flow around the body and associated pressure field, a novel multi overlapping grid method is implemented. Each grid having its own free surface, a two-way communication is ensured between the problem in the body vicinity and the larger scale wave propagation problem. Pressure field and nonlinear loads on the structure are computed by solving a boundary value problem on the time derivative of the potential.
     The stability and convergence properties of the solver are studied basing on extensive tests with standing waves of large to extreme wave steepness, up to $H/\lambda=0.2$ ($H$ is the crest-to-trough wave height and $\lambda$ the wavelength). Ranges of optimal time and spatial discretizations are determined and high-order convergence properties are verified, first without using any filter. For cases with either high level of nonlinearity or long simulation duration, the use of mild Savitzky-Golay filters is  shown to extend the range of applicability of the model.
    Then, the NWT is tested against two wave flume experiments, analyzing forces on bodies in various wave conditions.  First, nonlinear components of the vertical force acting on a small horizontal circular cylinder with low submergence below the mean water level are shown to be accurately simulated up to the third order in wave steepness. The second case is a dedicated experiment with a floating barge of rectangular cross-section. This very challenging case (body with sharp corners in large waves) allows to examine the behavior of the model in situations at and beyond the limits of its formal application domain. Though effects associated with viscosity and flow separation manifest experimentally, the NWT proves able to capture the main features of the wave-structure interaction and associated loads. \par
\end{abstract}

\paragraph{keywords:}
    Fully nonlinear waves,
    Numerical Wave Tank ,
    Harmonic Polynomial Cell,
    Boundary Value Problem,
    Immersed Boundary Method,
    Immersed overlapping grids

\section{Introduction and scope of the study \label{intro}}
%
To model the propagation of oceanic and coastal water waves of high steepness and their interaction with offshore structures fast and accurate numerical models are needed by scientists and engineers. Questions related to an accurate estimation of wave loads and structure dynamics in moderate to severe sea state conditions are of utmost importance for optimizing the design of such structures at sea and, for instance, assessing their behavior in extreme storm conditions and/or for fatigue analysis in the life-cycle of the structure. A wide range of models have been developed to predict wave fields and hydrodynamic loads at any scale, from the simple linear potential boundary element method to complex Computational Fluid Dynamics (CFD) codes directly solving the Navier-Stokes equations.

In the last decade, the use of CFD code has become increasingly popular in the scientific community, see for instance the applications of industrial or research codes like OpenFOAM\textregistered\ 
\citep{jacobsenFuhrmanFredsoe2012,hu2016numerical,higuera2013realistic,windt2019assessment}, %
STAR-CCM+ 
\citep{oggiano2017reproduction%
,jiao2020cfd}, 
ANSYS-FLUENT 
\citep{kim2016numerical,%
feng2019generation}%
, and REEF3D \citep{bihs2016new}%
, just to mention few of them.

    Although these CFD models are well adapted to solve the wave-structure interaction at small scale, in particular when complex physical processes such as wave breaking, formation of jets, air entrapment occur, their use is still mostly restricted to a limited number of  cases by the computational cost when targeting applications on ocean domains where the accurate representation of wave propagation is of first importance. Limitations associated with the employed numerical methods (\eg numerical diffusion and difficulty to resolve the dynamics of the free surface) are other reasons which still hinder the applicability of such models to large scale wave propagation problems. Their performance can be greatly increased by carefully selecting the different numerical schemes and parameters, to reduce the computational cost at the expense of some accuracy \citep{larsen_performance_2019}.
    
    On the other hand, models based on a potential flow approach (\ie neglecting viscous effects and assuming irrotational flow) are widely used to describe the dynamics of wave-structure interaction flows \citep[see \eg the review by][]{tanizawa2000state}. %
Among potential models, simplified linear versions are often used in the engineering field due to their low computational cost, allowing to capture the main effects on a wide range of parameters at a contained CPU cost. WAMIT \citep{lee1995wamit}%
, ANSYS-AQWA \citep{ansys2013aqwa} %
 or Nemoh \citep{babarit2015theoretical} are examples of such linear models. However, the linear assumption is often used outside its prescribed validity domain, for example in extreme cases where the allegedly small parameters, usually the wave steepness, becomes large. In those conditions many aspects of the dynamics of both the incident waves and the wave-body interaction are not properly modeled. 

Efforts have been done to extend these potential linear models to weakly nonlinear conditions at second order, mostly by adding terms like the Froude-Krylov forces or the complete Quadratic Transfer Functions, see \eg \citet{pinkster1980low}  or more recently \citet{philippe:hal-01198807}. 
However in some cases, mostly extreme wave conditions, even third order effects can have a significant effect on the wave load \citep[see \eg][]{fedele2017sinking}. Both the nonlinear wave dynamics and the nonlinear wave-structure interaction need to be taken into account. In those cases, alternative approaches capturing higher-order or fully nonlinear effects are needed. Several developments have been made to compute such nonlinear effect exactly or at high orders in the case of periodic regular waves in uniform water depth. For example, models based on analytical theories such as the Stokes wave theory \citep[see \eg ][]{sobey1989variations} %
or the so-called stream function method \citep{dean1965stream, fenton1988numerical} can only be applied with constant or simple geometry of the sea floor. A review of methods to describe wave propagation in a potential flow framework is given by \citet{fenton1999numerical}. 

High order spectral (HOS) methods are also fast and accurate to compute the flow behavior even up to large wave steepness. Such methods, though very efficient in computing the wave elevation even for large domains, might be limited in terms of geometry (complex bodies or variable bathymetry). They are mostly applied for wave maker modeling \citep{ducrozet2012modified}, or to compute the incident wave field within a more complex method to resolve the wave-structure interactions. For example, in the SWENSE method \citep[see \eg][]{luquest2007simulation}, the diffracted field is computed separately with a Navier-Stokes based solver. 

In order to develop a versatile Numerical Wave Tank (NWT), with the possibility to include bodies or sea bed with complex geometries, a time domain resolution involving a mesh in the spatial coordinates appears to be practical at the cost of an increase in the computational load. 
The commonly used Boundary Element Method (BEM), in which the Laplace equation is projected onto the spatially discretized boundaries using the Green's identities, has been proven to be effective in both 2D and 3D cases. For example, \citet{grilli1997numerical} used a high-order BEM method to generate and absorb waves in 2D. This model was extended to 3D by \citet{grilli_fully_2001}. For an overview of work on the BEM methods up to the end of the 20\textsuperscript{th} century, the reader is referred to \citet{kim1999recent}. More recently, \citet{guerber2012fully} and \citet{dombre_3d_2019} presented in great detail and implemented a complete NWT, in 2D and 3D respectively. Note that with the BEM schemes, a special attention must be given to the treatment of sharp corners \citep{hague_multiple_2009}. 

Another type of time domain wave simulators is volume field solvers. 
With the cost of increasing the number of unknowns, the resulting matrix is mostly sparse, allowing the use of efficient solvers. 
Both potential and Navier-Stokes solvers can be found in this category. Notable works on NWT based on the finite difference method (FDM) are given by \citet{tavassoli2001interactions}, or \citet{engsig-karup_efficient_2009} (OceanWave3D code). Finite element methods (FEM) were also successfully applied to both the potential problem \citep[\eg][]{ma_quasi_2006, yan_numerical_2007, engsig-karup_stabilised_2016, engsig-karup_mixed_2019} and the Navier-Stokes equations \citep[\eg][]{wu_simulation_2013}.

A different potential model that solves for the volume field was recently proposed by \citet{HPC:shao2012towards,HPC:shao2014fully,HPC:shao2014harmonic}  and tested against several methods including the BEM, FDM and FEM. This innovative technique, called the "harmonic polynomial cell" (HPC) method, was proven to be promising both in 2D and 3D \citep{HPC:shao2014fully,HPC:hanssen2015harmonic,HPC:hanssen2017free}. Although relatively new, this method was used to study a relatively important range of flows and phenomena: from a closed flexible fish cage \citep{strand2019linear} to hydrodynamics lifting problems \citep{liang2015application}. It was also extended to solve the Poisson equation by \citet{HPC:bardazzi2015generalized}. The numerical aspects of the method were studied in details by \citet{HPC:ma2017local} and applied as a 2D NWT by \citet{hpc:zhu2017improved}.

The treatment of the free surface conditions and body boundary condition can be done in several ways. A classical and straightforward approach is to use a grid that conforms the boundary shape. With this method, the boundary node values are explicitly enforced in the linear system. The drawback of this method is that the grid needs to be deformed at each time step so as to match the free surface. A lot of solutions have been successfully applied to tackle this issue. For instance, \citet{ma_quasi_2006} used a Quasi Arbitrary Lagrangian-Eulerian (ALE) method combined with the FEM spatial discretization to prevent the mesh to have to be regenerated at each time step. \citet{yan_numerical_2007} extended the mesh conformation technique to include a freely floating body. With this same FEM discretization, \citet{wu_simulation_2013} used in addition an hybrid Cartesian immersed boundary method for the body boundary condition. 

In this work, a NWT based on the HPC method is developed, its convergence assessed, and tested against both numerical and experimental data. Relaxation zones are used to generate and absorb waves in a similar manner as in the OpenFOAM\textregistered \ toolbox waveFoam \citep{jacobsenFuhrmanFredsoe2012}: the values and positions of the free surface nodes are imposed from a stream-function theory over a given distance. The free surface is tracked in a semi-Lagrangian way following \citet{HPC:hanssen2017free} whereas for the solid bodies, an additional grid fitted to the boundary is defined using the advances presented in \citet{HPC:ma2017local}. In order for the body to pierce the free surface, an additional free surface marker list is defined in the body fitted grid. "External" and "internal" curves -- \ie free surface boundaries evolving in the background and body fitted grids respectively -- also overlap each other and communicate through relaxation zones. In order to tackle the singular nodes, a null flow flux method is imposed on the sharp corners which lay on a Neumann body boundary condition in a similar manner as in \citet{hpc:zhu2017improved}, but applied at external body corners (described in \cref{ssec:singularNodes}).

The remainder of this article is organized as follows. In \cref{sec:math_pb} the mathematical formulation of the potential flow problem is recalled. The numerical methods based on the HPC approach are presented in \cref{sec:HPC}, with particular attention devoted to the treatment of the free surface. In \cref{sec:standing_w}, a convergence study is performed on a freely evolving standing wave case compared to a highly accurate numerical solution. Numerous tests are performed to determine the optimal ranges of discretization parameters, first without using any filter. The effects associated with mild filters are then analyzed. The case of a standing wave of near maximum amplitude is run to confirm the nonlinear capabilities of the model. Then, a novel fitted mesh overlapping grid method is described and implemented  in \cref{sec:theory_fixedBody}. This double mesh strategy is tested on two selected cases in \cref{sec:valid}. The first one is an horizontal fixed circular cylinder, completely immersed although close to the free surface, from \citet{chaplin1984nonlinear} and the second one involves a free surface piercing rectangular barge. That second experiment is an original campaign done during this study. Results of the present NWT on the two test cases are compared with experimental data and numerical results from the literature. In \cref{sec:conclusions}, the main findings from this study and research outlook are summarized and discussed.

\section{Fully nonlinear potential flow modeling approach} \label{sec:math_pb}
Three main assumptions are used in the potential model: i. we consider a fluid of constant and homogeneous density $\rho$ (incompressible flow); ii. the flow is assumed to be irrotational, implying that $\nabla\times{\tI{v}}=0$, where $\tI{v}(x,y,z,t)$ denotes the velocity field; and iii. viscous effects are neglected (ideal fluid). Here, $(x,y)$ denote horizontal coordinates, $z$ the vertical coordinate on a vertical axis pointing upwards, and $t$ the time. The gradient operator is defined as $\nabla f \equiv (f_x,f_y,f_z)^T$, where subscripts denote partial derivatives (\eg $f_x=\frac{\partial f}{\partial x}$).

The complete description of the velocity field $\tI{v}$ can thus be reduced to the knowledge of the potential scalar field $\phi(x,y,z,t)$, such that:
$\tI{v}=\nabla \phi$. 
Due to the incompressibility of the flow, the potential $\phi$ satisfies the Laplace equation inside the fluid domain:
\begin{equation}
    \nabla^2 \phi  =  0 , \hspace{2cm} 	 -h(x,y) \leq z \leq \eta(x,y,t) \label{eq:Laplace}
\end{equation}
where $\eta(x,y,t)$ is the free surface elevation and $h(x,y)$ the water depth relative to the still water level (SWL).
In order to solve this equation, boundary conditions need to be considered. On the time varying free surface $z=\eta(x,y,t)$, the Kinematic Free Surface Boundary Condition (KFSBC) and the Dynamic Free Surface Boundary Condition (DFSBC) apply:

\begin{align}
    \label{eq:KFSBC} \eta_{t}+\nabla_H \eta \cdot \nabla_H \phi-\phi_z &=0 \quad \text{on } z = \eta(x,y,t),\\
    \label{eq:DFSBC} \phi_{t}+\frac{1}{2}\left(\nabla_H \phi\right)^2+g\eta &=0 \quad \text{on } z = \eta(x,y,t),
\end{align}
where $\nabla_H f \equiv (f_x,f_y)^T$ denotes the horizontal gradient operator and $g$ the acceleration due to gravity. At the bottom (impermeable and fixed in time), the Bottom Boundary Condition (BBC) reads:

\begin{equation}
    \nabla_H h \cdot \nabla_H \phi+\phi_z =0 \quad \text{on } z = -h(x,y). \label{eq:BBC}
\end{equation}
On the body surface, the slip boundary condition expresses that the velocity component of the flow normal to the body face equals the normal component of the body velocity. Here, we restrict our attention to fixed bodies, thus, denoting $\tI{n}$ the unit vector normal to the body boundary, this condition reduces to:

\begin{equation}
    \frac{\partial \phi}{\partial \tI{n}} = \nabla \phi \cdot \tI{n} =0 \quad \text{on the body}. \label{eq:bodyBC}
\end{equation}

Note that, using the free surface velocity potential and the vertical component of the velocity at the free surface, defined respectively as:

\begin{eqnarray}
	\tilde{\phi}(x,y,t) & = & \phi(x,y,\eta(x,y,t),t) \label{eq:fspot}\\
	\tilde{w}(x,y,t) & = & \frac{\partial \phi}{\partial z}(x,y,\eta(x,y,t),t) \label{eq:fsw}
\end{eqnarray}
the KFSBC and the DFSBC can be reformulated following~\citet{zakharov1968} as:

\begin{align}
\eta_{t} & =  -\nabla_H \eta \cdot \nabla_H \tilde{\phi}+\tilde{w}(1+(\nabla_H \eta)^2)  \label{eq:zakharov1}\\
\tilde{\phi}_t & =  -g\eta-\frac{1}{2}(\nabla_H \tilde{\phi})^2+\frac{1}{2}\tilde{w}^2(1+(\nabla_H \eta)^2) \label{eq:zakharov2}
\end{align}
It can be noted that the Laplace \cref{eq:Laplace}, the BBC~(\cref{eq:BBC}) and the body BC~(\cref{eq:bodyBC}) are all linear equations. Thus, the nonlinearity of the problem originates uniquely from the free surface boundary conditions (\cref{eq:KFSBC,eq:DFSBC}) or (\cref{eq:zakharov1,eq:zakharov2}).
 A classical approach at that point is to assume that surface waves are of small amplitude relative to their wavelength%
, so that these boundary conditions can be linearized and applied at the SWL (\ie at $z=0$). Here, we intend to retain full nonlinearity of wave motion by considering the complete conditions~(\cref{eq:zakharov1,eq:zakharov2}). These two equations are used to compute the time evolution of the free surface elevation $\eta$ and the free surface potential $\tilde{\phi}$. This requires obtaining $\tilde{w}$ from $(\eta,\tilde{\phi})$, a problem usually referred to as Dirichlet-to-Neumann (DtN) problem.

For given values of $(\eta,\tilde{\phi})$, the DtN problem is here solved by solving a BVP problem in the fluid domain on the wave potential $\phi(x,y,z,t)$, composed of the Laplace equation~(\cref{eq:Laplace}), the BBC~(\cref{eq:BBC}), the body BC~(\cref{eq:bodyBC}), the imposed value $\phi(z=\eta)=\tilde \phi$ (Dirichlet condition) on the free surface $z=\eta$, supplemented with boundary conditions on lateral boundaries of the domain (of \eg Dirichlet, Neumann, etc. type). The numerical methods to solve the BVP are presented in the next section.

\section{The Harmonic Polynomial Cell method (HPC) with immersed free surface} \label{sec:HPC}
\subsection{General principle of the HPC method} \label{ssec:HPC_0}
In order to solve the above mentioned BVP at a given time, the HPC method introduced by~\citet{HPC:shao2012towards} is used. It is briefly described here, and more details can be found in~\citet{HPC:shao2014harmonic},~\citet{HPC:hanssen2015harmonic,HPC:hanssen2017free},~\citet{HPC:hanssen2019non} and~\citet{HPC:ma2017local}. In this work, the HPC approach is implemented and tested in 2 spatial dimensions, \ie in the vertical plane $(x,z)$, for a wide range of parameters.

\begin{figure}[htbp!]
	\begin{center}
        \includegraphics{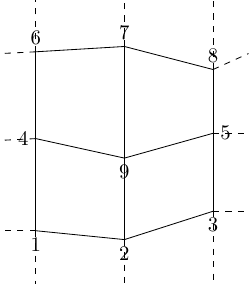}
	\caption{Definition sketch of 9-node macro-cell used for the HPC method, with local numbering of the nodes}
	\label{sch:HPC_cell}
	\end{center}
\end{figure}

The fluid domain is discretized with overlapping macro-cells which are composed of 9 nodes in 2 dimensions. Those macro-cells are obtained by assembling four adjacent quadrilateral cells on an underlying quadrangular mesh. The four cells of a macro-cell share a same vertex node, called the "central node" or "center" of the macro-cell. A typical macro-cell is schematically shown in \cref{sch:HPC_cell}, with the corresponding local index numbers of the 9 nodes. With this convention, any node with global index $n$ has the local index "9" in the considered macro-cell  and is considered as an interior fluid point, whereas for example node with local index "4" can either be a fluid point or a point lying an a boundary. %
In the case node ``4'' is also an interior fluid point, it defines a new macro-cell overlapping the original one: nodes denoted 6,7,8,9,2 and 1 in \cref{sch:HPC_cell} also belong to this macro-cell centered by ``4''.

In each macro-cell, the velocity potential is approximated as a weighted sum of the 8 first harmonic polynomials (HP), the latter being fundamental polynomial solutions of the Laplace \cref{eq:Laplace}. 
A discussion about which of the HP are to be chosen is given in \citet{HPC:ma2017local}. Here, we follow \citet{HPC:shao2012towards}, and select all polynomials of order 0 to 3 plus one fourth-order polynomial, namely: $f_1(\tI{x})=1$, $f_2(\tI{x})=x$, $f_3(\tI{x})=z$, $f_4(\tI{x})=xz$, $f_5(\tI{x})=x^2-z^2$, $f_6(\tI{x})=x^3-3xz^2$, $f_7(\tI{x})=-z^3+3x^2z$ and $f_8(\tI{x})=x^4-6x^2z^2 +z^4$. Here, $\tI{x}=(x,z)$ represents the spatial coordinates. Thereafter, we define $\tI{\bar{x}} = \tI{x}-\tI{x}_9$ the same spatial coordinate in the local reference frame of the macro-cell,  with $\tI{x}_9$ being the center node of the macro cell. %
From a given macro-cell, the potential can be approximated at a location $\tI{x}$ as:

\begin{equation}
	\phi(\tI x)=\sum_{j=1}^{8}b_j f_j(\tI{\bar{x}}) \label{eq:interp}
\end{equation} 
As every HP is a solution of the Laplace \cref{eq:Laplace} which is linear, any linear combination of them is also solution of this equation. Thus, the goal now becomes to match the local expressions (LE) given by \cref{eq:interp}%
: if this equation is satisfied at each node ($j=1,...,8$) of each macro-cell, a solution of the BVP is available everywhere. %
For that reason, macro-cells overlap each other. Note that this study deals with 2D problems, but the method can be extended to 3D cases as shown by~\citet{HPC:shao2014fully} considering cubic-like macro-cells with 27 nodes.

The first objective is to determine the vector of coefficients $b_j, j=1,...,8$ for the selected macro-cell. Recalling that \cref{eq:interp} should be verified at the location of each point of the macro-cell (with local index running from $1$ to $9$), this equation applied at the 8 neighboring nodes $(1-8)$ of the center yields:

\begin{equation}
    \phi_i=\phi(\tI{x}_i) =  \sum_{j=1}^{8} b_j f_j(\tI{\bar{x}}_i) \quad \text{for} \ i=1,...,8
\end{equation}
which represents, in vector notation, a relation between the vector of size 8 of the values of the potential $\phi_i$ at the outer nodes with the vector of size 8 of the $b_j$ coefficients.
The 8x8 local matrix linking these two vectors is denoted $\tII{C}$, and defined by  $C_{ij}=f_j(\tI{x}_i)$.
Note that $\tII{C}$ is defined geometrically, thus it only depends on the position of the outer nodes $i$ relatively to the position of the central node.
$\tII{C}$  can be inverted and its inverse is denoted $\tII{C}^{-1}$. The $b_j$ coefficients are then obtained for the given macro-cell as a function of the potentials at the 8 neighboring nodes of the central node of that macro-cell: 

\begin{equation}
    b_j = \sum_{i=1}^{8} C^{-1}_{ji}\phi_i \quad \text{for} \ j=1,...,8.
\end{equation}
Injecting this result into the interpolation \cref{eq:interp}, a relation is found providing an approximation for the potential at any point located inside the macro-cell using the values of the potential of the eight surrounding nodes of the central node: 

	\begin{equation}
		\phi(\tI x)= \sum_{i=1}^{8}  \left(\sum_{j=1}^{8} C^{-1}_{ji}  f_j(\tI{\bar{x}}) \right) \phi_i \label{eq:mainwithcij}
	\end{equation}

This equation will be referred to as local expression (LE) of the potential. It will be used to derive the boundary conditions equations and the fluid node equations that need to be solved in the BVP. Also note that this LE provides a really good interpolation function that can be used for every additional computation once the nodal values of the potential are known (\ie potential derivatives at the free surface or close to the body to compute the pressure field).
 
Note that the accuracy of LE depends only on the geometry: coordinates at which this equation is applied, shape of the macro-cell, etc. Those dependencies are investigated in details by~\citet{HPC:ma2017local}.
%
\subsection{Treatment of nodes inside the fluid domain} \label{ssec:HPC_1}
We first consider the general case of macro-cells whose central node is an interior node of the fluid domain. Applying the LE~(\cref{eq:mainwithcij}) at the central node yields a linear relation between the values of the potential at the nine nodes of this macro-cell: 

\begin{equation}
    \phi_9=\phi(\tI{x}_9)= \sum_{i=1}^{8}  \left(\sum_{j=1}^{8} C^{-1}_{ji}  f_j(\tI{ \bar{x}}_9) \right) \phi_i
    \label{eq:mainatx9}
\end{equation}
We may further simplify this equation by noting that, as $\tI{\bar{x}}_9=(0,0)$ in local coordinates, all $f_j(\tI{\bar{x}}_9)$ vanish, except $f_1(\tI{\bar{x}}_9)$ which is constant and equal to $1$. \Cref{eq:mainatx9} then simplifies to:

\begin{equation}
    \phi_9= \sum_{i=1}^{8}  C^{-1}_{1i} \phi_i
    \label{eq:mainatx9b}
\end{equation}
meaning that only the first row of the matrix $\tII{C}^{-1}$ is needed here.

In order to solve the global potential problem, \ie to find the nodal values of the potential at all grid points (whose total number is denoted $N$), a global linear system of equations is formed, with general form $\tII{A}.\tI{\phi}=\tI{B}$, or:

\begin{equation}
    \sum_{l=1}^{N} A_{kl}\phi_l=B_k \quad \text{for} \ k=1,...,N.
\end{equation}
where $k$ and $l$ are global indexes of the nodes.
For each interior node in the fluid domain, with global index $k$ and associated macro-cell, an equation of the form of \cref{eq:mainatx9b} allows to fill a row of the global matrix $\tII{A}$. This row $k$ of the matrix involves only the considered node and its 8 neighboring nodes, making the matrix $\tII{A}$ very sparse (at most 9 non-zero elements out of $N$ terms). Moreover, the corresponding right-hand-side (RHS) term $B_k$ is null. Note that all the 8 neighboring nodes of the macro-cell associated with center $k$ should also have a dedicated equation in the global matrix in order to close the system. 
%
\subsection{Nodes where a Dirichlet or Neumann boundary condition is imposed
\label{sec:sub:dirandneucond}}
If a Dirichlet boundary condition with value $\phi_D$ of the potential has to be imposed at the node of global index $k$, the corresponding equation is simply $\phi_k=\phi_D$, so that only the diagonal element of the global matrix is non-null and equal to 1 for the corresponding row $k$: $A_{kl}=\delta_{kl} \ \ \forall l \in [1,N]$. The corresponding term on the RHS is set to $B_k=\phi_D$.

If a Neumann condition has to imposed at a given node of global index $k$, the relation set in the global matrix is found through the spatial derivation along the imposed normal $\tI{n}$ of the LE~(\cref{eq:mainwithcij}) of any macro-cell on which $k$ appears. In practice, the macro-cell whose center is the closest to the node $k$ is chosen, and we then use: 

\begin{equation}
	\nabla \phi(\tI{ x}_k) \cdot \tI{n} = \sum_{i=1}^{8}  \left(\sum_{j=1}^{8} C^{-1}_{ji}  \nabla f_j(\tI{ \bar{x}}_k) \cdot \tI{n} \right) \phi_i \label{eq:mainderivated}
\end{equation}

Thus, a relation is set in the row $k$ of the global matrix to enforce the value of $B_k=\nabla \phi(\tI{x}_k) \cdot \tI{n}$ at position $\tI{x}_k$. In that case, a maximum of 8 non-zero values appear in this row on the global matrix as the potential of the central node of the macro-cell does not intervene here.

\subsection{Treatment of the free surface} 
As already mentioned, in order to solve the BVP at a given time-step, the system of equations needs to be closed, meaning that each neighbor of a node in the fluid domain should have a dedicated equation. We consider now the case of nodes lying on or in the vicinity of the (time varying) free surface. The free surface potential should be involved here, either directly at a node fitted to the free surface through a Dirichlet condition described in the previous sub-section, or through alternative techniques.

For instance, an Immersed Boundary Method (IBM) was first suggested in the HPC framework by \citet{HPC:hanssen2015harmonic} to tackle body boundary conditions. 
More recently,~\citet{HPC:ma2017local} compared a modified version of the IBM with two different multi-grid (MG) approaches (fitted or combined with an IBM) for both body and free surface boundary conditions. \citet{HPC:hanssen2017free} and \citet{HPC:hanssen2017wave} also made in-depth comparisons of the MG and IB approaches, focusing on the free surface tracking. Both methods showed promising results. \citet{hpc:zhu2017improved} introduced a similar yet slightly different IB approach with one or two ghost node layers, then realized a comparison between this IB approach and the original fitted mesh approach.
In the present work, the IBM was chosen for the treatment of the free surface, though the fitting mesh method is shortly described thereafter.
\subsubsection{Fitted mesh approach for the free surface \label{sec:sub:sub:oldfittedMesh}}
The first possibility is to fit the mesh to the actual free surface position at any time when the BVP has to solved. The mesh is deformed so that the upper node at any abscissa always lies on the free surface.
That way, the computational domain is completely closed and the free surface potential is simply enforced as a Dirichlet boundary condition at the correct position $z=\eta$ as explained in \cref{sec:sub:dirandneucond}. %
With this approach, the algorithm, given the boundary values at the considered time, can be summarized as:
\begin{itemize}
	\item Deform the mesh to fit the current free surface elevation,
	\item Build and then invert the local geometric matrices $\tII{C}$,
    \item Fill the global matrix $\tII{A}$ and RHS $\tI{B}$, using the corresponding Dirichlet conditions at nodes lying on the free surface,
	\item Invert the global problem to obtain the potential everywhere
\end{itemize}
Recently, \citet{HPC:ma2017local} pointed out that the HPC method efficiency (in terms of accuracy and convergence rate) is greatly improved when a fixed mesh of perfectly-squared cells is used. In this work, the negative effects of a deforming mesh outlined in the previous subsection were also encountered. Especially, for some particular cell shapes, a high increase of the local condition number was observed, leading to difficulty of matrix inversion and important errors on the approximated potential. %
As a consequence, results were highly dependent on the mesh deformation method employed, especially in the vicinity of a fixed fully-immersed body.
\subsubsection{Immersed free surface approach \label{sec:sub:ifs}}
In order to work with regular fixed grids, an IBM technique was developed and implemented to describe the free surface dynamics. \citet{HPC:hanssen2015harmonic} introduced a first version of this method applied on the boundaries of a moving body. This method was recently extended to the free surface and compared to a fitted MG method by \citet{HPC:ma2017local} and \citet{HPC:hanssen2017free}. In the current work, a semi-Lagrangian IB method introduced by \citet{HPC:hanssen2017free} is chosen.

\begin{figure}[htbp!]
	\begin{center}
        \includegraphics{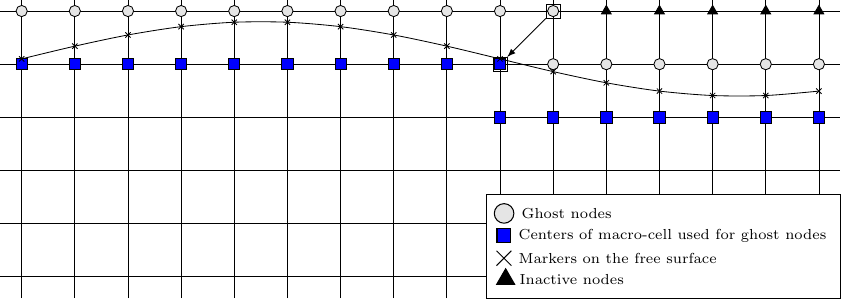}
	\caption{Schematic representation of the immersed free surface in a fixed grid}
	\label{sch:ifreesurface}
	\end{center}
\end{figure}

In this method, the free surface is discretized with markers, evenly spaced and positioned at each vertical intersection with the background fixed grid, as shown in \cref{sch:ifreesurface}. Those markers are semi-Lagrangian in such a way that they are only allowed to move vertically, following \cref{eq:zakharov1,eq:zakharov2}. 

At a given time, every node located below the free surface (\ie below a marker) is considered as a node in the fluid domain ("fluid" node), and defines a macro-cell with its 8 neighbors.
The global matrix is classically filled with the local expression~(\cref{eq:mainwithcij}) at those nodes. 
As a consequence, in order to close the system, each neighbor of a node just above the free surface must also have a dedicated equation in the global matrix. These neighbors, represented with grey circles on \cref{sch:ifreesurface}, are denoted as "ghost" nodes.
The chosen equation to close the system at a node of this type is the local expression~(\cref{eq:mainwithcij}) applied at the marker position in a given macro-cell: 

\begin{equation}
    \phi_m=\sum_{i=1}^8 \left(\sum_{j=1}^8 C_{ji}^{-1}f_j(\tI{\bar {x}}_m)\right) \phi_i
    \label{eq:markerinterp}
\end{equation}
where $\bar{\tI{x}}_m = (x_m,\eta(x_m))-  \tI{x}_c$ is the position of the marker in the macro-cell's reference frame ($\tI{x}_c$ is the global position of the center node of the chosen macro-cell) and $\phi_m$ its potential (known at this stage).
This ensures that the potential at the free surface point is equal to the potential at the position of the marker from the interpolation equation. In other words, if one wants to interpolate the computed field $\phi$ at the particular location of the marker $\tI{x}_m$, the results should be consistent and yield the potential $\phi_m$.

Note that this \cref{eq:markerinterp} is cell dependent (through $C_{ji}^{-1}$, the involved $\phi_i$ and the position of the %
center node $\tI{x}_c$)%
, but also depends on the chosen marker (through $\tI{x}_m$ and $\phi_m$).
The only mathematical restriction on the choice the macro-cell to consider is that the ghost point potential should intervene as one of the $\phi_i$ in order to impose the needed constraint at this point.

An important note is that the latter \cref{eq:markerinterp} is not dependent on the ghost point in any fashion.
This implies that if the same couple (marker, macro-cell) is chosen to close the system at two different ghost points, the global matrix will have two strictly identical rows. Its inversion would thus not be possible. %
Particularly, two vertically aligned ghost points cannot use the same macro-cell equation at the same marker position. 
Here stands the difference between the IB method of \citet{HPC:hanssen2017wave}, \citet{HPC:ma2017local} and the one chosen by \citet{hpc:zhu2017improved}.
\citet{hpc:zhu2017improved} decided to only impose the marker potential once in the first layer (or two first layers) and to constraint the upper potentials to an arbitrary value (in practice if the point is not used directly, the potential is set to the first point below which potential is used). 
The method used during this work is closer to the one by \citet{HPC:hanssen2017wave} and \citet{HPC:ma2017local}: if a node needs a constraint but does not have a marker directly underneath (case of two ghost points vertically aligned), the ghost point on the top should invoke the local expression of the macro-cell centered on the closest fluid point instead of the cell centered on the vertically aligned fluid point (case indicated by an arrow in \cref{sch:ifreesurface}). 
With that method, in such a situation, the potential of %
the selected marker -- \ie vertically aligned with the two ghost points -- %
is imposed twice in two different adjacent macro-cells.
A comparison between those two methods had not been conducted and would be of great interest.

Regardless of the kind of IB method used, the main goal is achieved: it is not needed to deform the mesh in time.
As a consequence, the computation and inversion of the local (geometric) matrices is only done once, at the beginning of the computation.
However, a step of identification of the type of each node, which was proven to be time consuming, is needed instead. Note that this identification algorithm could be greatly improved and is relatively slow in its current implementation. 
The general algorithm at one time step becomes:
\begin{itemize}
	\item Identify nodes inside the fluid domain,
	\item Identify ghost nodes needed to close the system, associated markers and macro-cells,
  \item Fill global matrix $\tII{A}$ and RHS $\tI{B}$,
	\item Invert global problem to obtain the potential everywhere,
\end{itemize}
	
\subsection{Linear solver and advance in time} \label{ssec:linsol_time}
To solve the global linear sparse system of equations, an iterative GMRES solver, based on Arnoldi inversion, was used for all computations. The base solver was developed by \citet{gmres:saad2003iterative} for sparse matrix (SPARSEKIT library), and includes an incomplete LU factorization preconditionner. In the current implementation, a modified version of the latter is used with the improvement proposed by \citet{baker_simple_2009}. Except for stalling during the study of a standing wave at very long time, this solver was proven to be robust.
Improvements of the construction step of the global matrix could further be made in order to increase the efficiency of its inversion. Also, the initial guess in the GMRES solver could also be improved taking advantages of the already computed potential values.
The number of inner iterations of the GMRES algorithm was chosen as $m \in [30, 60]$ and the iterative solution is considered converged when the residual is lower that $5.10^{-9}$.

Marching in time thanks to \cref{eq:zakharov1,eq:zakharov2} yields the free surface elevation and the free surface potential at the next time step.
Note that the steps of computing the RHS terms of these equations are straightforward for most terms directly from the local expression~(\cref{eq:mainwithcij}) of the closest macro-cell. 
In addition, the spatial derivative of $\eta$ is computed with a finite difference method. A centered scheme of order 4 is chosen for this work with the objective to maintain the theoretical order 4 of spatial convergence provided by the HPC method.

In order to integrate \cref{eq:zakharov1,eq:zakharov2}, the classical four-step explicit Runge-Kutta method of order 4 (RK4) was selected as time-marching algorithm.  During a given simulation, the time step $(\delta t)$ was chosen to remain constant. Its value is made nondimensional by considering the Courant-Friedrichs-Lewy (CFL) number $C_o$ based on the phase velocity $C=\lambda/T$, where $\lambda$ is the wavelength and $T$ the wave period:

\begin{equation}
    C_o=\frac{C \delta t}{\delta x}=\frac{\lambda/\delta x}{T/\delta t}
    \label{eq:CFL}
\end{equation}
The CFL number thus corresponds to the ratio of the number of spatial grid-steps per wavelength $(N_x=\lambda/\delta x)$ divided by the number of time-steps per wave period $(N_t=T/\delta t)$, \ie $C_o=N_x/N_t$.

\subsection{Computation of the time derivative of the potential}\label{ssec:potentialderivative}
The pressure inside the fluid domain is obtained from the Bernoulli equation: 

\begin{equation}
    p(x,z,t)=-\rho \left( \dfrac{\partial \phi}{\partial t} + \dfrac{1}{2}(\nabla \phi)^2 +  gz \right) \label{eq:Bernoulli}
\end{equation}
This equation is used in every potential based NWT in order to obtain the wave loads. For this reason, the knowledge of the time derivative of the potential is required. %
In a first attempt, the time derivative of the potential was estimated using a backward finite difference scheme. However, this method is not well suited when important variations of the potential are at play. Moreover, in the case of the IB method, it is not possible to obtain the value of the pressure at a point that was previously above the free surface, and thus for which a time derivative of the potential cannot be computed by the finite difference scheme.

A fairly accurate method is to introduce the (Eulerian) time derivative of the potential as a new variable $\phi_t=\dfrac{\partial \phi}{\partial t}$ and to solve a similar BVP as described previously on this newly defined variable, noting that $\phi_t$ has to satisfy the same Laplace equation as $\phi$ in the fluid domain. This method, first used by \citet{cointe_nonlinear_1991,tanizawa_long_1996}, has been recently applied by \eg  \citet{guerber_modelisation_2011} in the BEM framework or by \citet{HPC:ma2017local} in the HPC method.

Note that the local macro-cell matrices and coefficients, which are only geometrically dependent, do not change. 
In the different expressions presented above that are used to fill the global matrix, the coefficients linking the different potentials are not time dependent. Thus, the matrix to invert is exactly the same for the $\phi_t$ field and the $\phi$ field.
However, the boundary conditions on $\phi$ and $\phi_t$ might differ, leading to different RHS. This is the case only when the RHS on $\phi$ is different from zero, as for example, for free surface related closure points. 

Remember that the equations at a (non-moving) Neumann condition and at a point inside the fluid domain yield a zero value in the RHS, and thus the equations at those points are exactly the same for the potential variable and for its derivative. 
At a (non-moving) Dirichlet boundary condition, one would simply impose $\phi_t=0$ instead of $\phi =\phi_D$. At the IB ghost points (\ie to enforce the free surface condition), the $\phi_t$ is imposed to match the derivative of the potential with respect to time, known at the marker positions from \cref{eq:zakharov2}.  
Even though the global matrices are exactly the same, the RHS being different and the chosen resolution method being iterative (GMRES solver), the easiest way is just to solve twice the almost same problem. A more clever way could maybe be investigated by taking advantage of the previous inversion, but this is left for future work. 

\section{Validation and convergence study on a nonlinear standing wave \label{sec:standing_w}}
%

\subsection{Presentation of the test-case \label{ssec:standing_w_intro}}
\begin{figure}[htbp]
    \centering
    \includegraphics{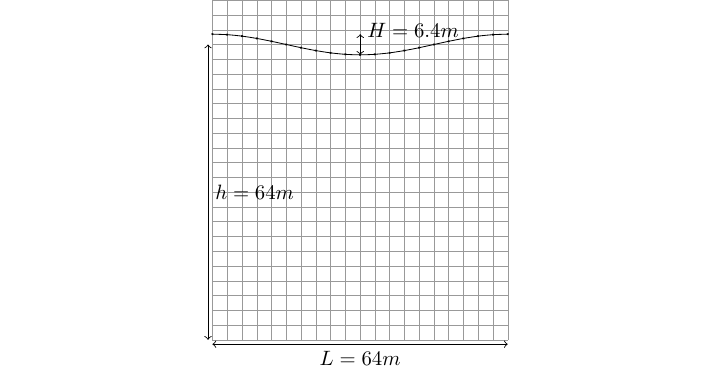}
    \caption{Schematic representation of the nonlinear standing wave with steepness $H/\lambda=10\%$ at $t=0$. Mesh grid represented with $N_x=20$.}
    \label{sch:standingwave}
\end{figure}

The first case consists in simulating a nonlinear standing wave in a domain of uniform water depth $h$ whose extent is equal to one wavelength $\lambda$. This case is quite demanding as the wave height $H$ (difference between the maximum and minimum values of free surface elevation at antinode locations) is fixed by choosing a large wave steepness $H/\lambda=10\%$ (or $kH/2=\pi/10\approx 0.314$). We also choose to work in deep water conditions by selecting $h=\lambda=64$~m (or $kh=2\pi\approx 6.28$). The water domain at rest is thus of square shape in the $(x,z)$ plane, as illustrated in \cref{sch:standingwave}.

Initial elevations of free surface $\eta(x,t=0)$ are computed from the numerical method proposed by \citet{tsai1994numerical}. The initial phase is chosen such that the imposed potential field is null at $t=0$ at any point in the water domain. This initial state corresponds to a maximum wave elevation at the beginning and the end of the domain ($x/\lambda=0$ and $1$), and a minimum wave elevation at the center point of the domain ($x/\lambda=0.5$), these three locations being antinodes of the standing wave.

Wall boundary conditions are enforced as null Neumann conditions, \ie $\nabla \phi \cdot \tI{n} = 0$ on the three wet walls, where $\tI{n}$ is the considered wall normal vector.%

The wave is freely evolving under the effect of gravity: in theory one should observe a fully periodic motion without any damping as the viscosity is neglected. At each time step, a spatial $L_2(\eta)$ error on $\eta$ is computed relative to the theoretical solution of \citet{tsai1994numerical} (denoted $\eta^{th}$ hereafter), and normalized with the wave height:

\begin{equation}
    L_{2}(\eta,t)=\dfrac{1}{H}\sqrt{\frac{1}{n_p}\sum_{i=1}^{n_p}{\left(\eta(x_i,t)-\eta^{th}(x_i,t)\right)^2}} 
\end{equation}
where $i$ represents the index of a point on the free surface and $n_p$ the total number of points on the free surface.

\subsection{Evolution of \texorpdfstring{$L_2(\eta)$}{L2n} error with space and time discretizations \label{ssec:standing_w_error}}
\begin{figure}[htbp!]
\centering
    \includegraphics{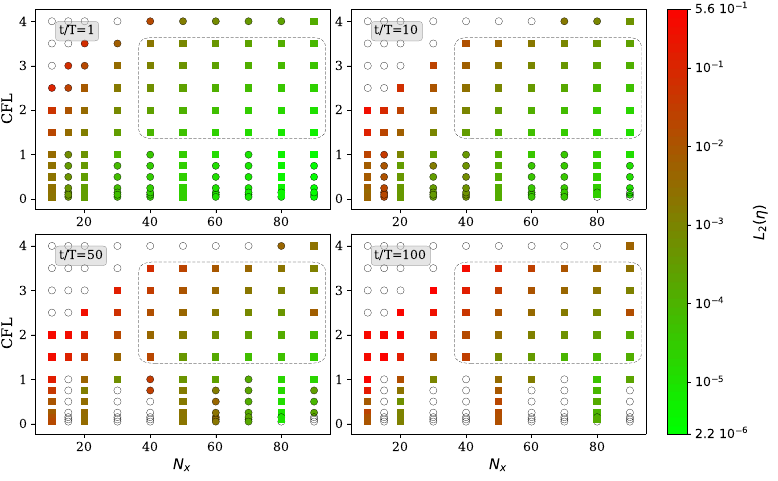}
    \caption{$L_2$ error on $\eta$ on the nonlinear standing wave case at four time instants ($t/T=1$, $10$, $50$ and $100$) as a function of the spatial and temporal discretizations. The color scale indicates the $L_2(\eta)$ error respective to the theoretical solution by \citet{tsai1994numerical}. See text for explanations on the significance of the markers shapes.}
    \label{fig:ErrorTotStanding}
\end{figure}

The result of this $L_2(\eta)$-error is represented as a color map at four different times $t/T =$ 1, 10, 50 and 100 in \cref{fig:ErrorTotStanding} as a function of the number of nodes per wavelength ($N_x=\lambda/\delta x$, where $\delta x$ is the spatial step-size) and the CFL number $C_o$ (defined in \cref{ssec:linsol_time}). Wide ranges of the two discretization parameters are explored, namely $N_x\in[10, 90]$ and $C_o$ $\in[0.05, 4.0]$.  Simulations that ran till the end of the requested duration of $100T$ are represented with coloured squares. A circle is chosen as a marker when the computation breaks down before the end of that duration. Nonetheless the markers are colored if the computation did not yet diverge at the time instant shown on the corresponding panel.

Results of \cref{fig:ErrorTotStanding} show that a large number of simulations were completed over this rather long physical time of $100T$. In this \S, no filter were used, so some of the numerical simulations tend to be unstable for extreme values of the discretization parameters. 
For instance, when $C_o \le 1$, the computation is mostly unstable and breaks down: before 50$T$ when $N_x$ is small (\ie below 40) and between 50$T$ and 100$T$ when $N_x$ is larger. Note that $C_o=1$ corresponds to a time step ranging from $N_t = T/\delta t=10$ to $90$, for $N_x=10$ and $90$ respectively. This value of $C_o=1$, and associated time step, is the lower stable limit exhibited by these simulations.

On the other hand, when the $C_o$ is too high (\ie larger than 3.5) instabilities also occur almost at the beginning of the simulation ($t/T<10$), particularly when $N_x$ is small. 
For very small $N_x$ (in the range 10-15) and whatever the $C_o$, the computation tends to be unstable. This is probably due to the discretization of the immersed free surface being the same as the discretization of the background mesh. 
A coarse discretization of the free surface leads to an inaccurate computation of the spatial derivative of $\eta$: instabilities may then occur.

A suitable range of parameters is thus determined to avoid instabilities: $1.5 \le C_o \le 3.5$ and  $40 \le N_x \le 90$. This zone is represented in \cref{fig:ErrorTotStanding} as a rectangular box with a dashed contour. In that zone, all the computations ran with the requested time step over a duration of $100T$. Note that the CPU cost scales with $N_x^2 N_t \sim N_x^3 /C_{o}$ and thus the most expensive computations in this stable zone are approximately 25 times slower than the least expensive ones in the same zone.
Also note that the lowest error is almost systematically reached in this zone. The $L_2(\eta)$ error is as small as $2.10^{-6}$ after $1T$. After $100T$, the lowest error is approximately $10^{-4}$. Moreover, the evolution of the value of the error is qualitatively consistent with the mesh refinement and time refinement.

The stability was not assessed for finer mesh than $N_x=90$ points per wave length, due to increasing computation cost on one hand, and the fact that finer resolutions would lie out of the range of discretizations targeted for real-case applications. In addition, at long time, a discretization of $N_x= 90$ already exhibits behavior that does not match the expected convergence rate, as will be discussed hereafter in greater detail.

\subsection{Convergence with time discretization \label{ssec:standing_w_conv_t}}
\begin{figure}[htbp!]
	\centering
    \subfloat[$t/T=1$]{\label{fig:sub:convt1}\includegraphics{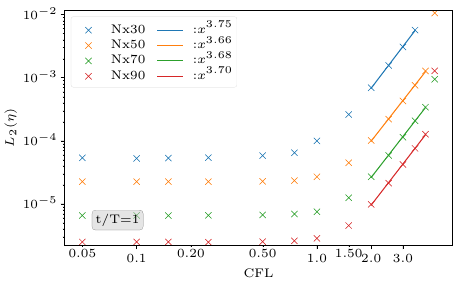}}
    \subfloat[$t/T=100$]{\label{fig:sub:convt100}\includegraphics{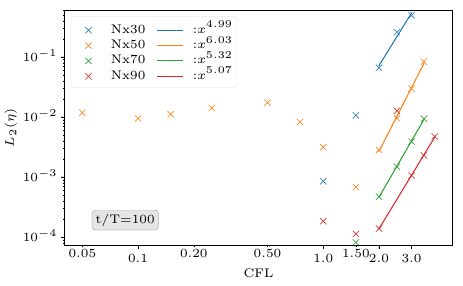}}
    \caption{Convergence of the $L_2$  error on $\eta$ (crosses) with respect to the temporal discretization at two different physical times: $t/T=1$ (left panel) and $t/T=100$ (right panel). The spatial discretization is fixed for a given line. Solid lines represent power regression of the error in the "linear range", the computed power is reported in the legend of the fitted straight lines.}
	\label{fig:convergencett1t100}
\end{figure}

In order to study the convergence of the method in a more quantitative manner, the $L_2(\eta)$ error is shown as a function of the $C_o$ number for different spatial discretizations $N_x$ at $t/T=1$ in \cref{fig:sub:convt1} and at $t/T=100$ in \cref{fig:sub:convt100}. A $C_o^\alpha$ regression line is computed and fitted on the linear convergence range of the log-log plot of the error. That will be called "linear range" for simplicity, though it corresponds to an algebraic rate of convergence of the error. Note that this linear range corresponds exactly to the zone in which the computations remain stable (with the exception of one particular point at $t/T=100$ and $N_x=90$ excluded from the determination of the convergence rate). At $t/T=1$, the minimum error is, as expected, obtained for small $C_o$ numbers and large $N_x$: $L_2(\eta) \sim 10^{-5}$ in the linear range and the minimal error reached is $2.10^{-6}$ for the finer discretization $N_x = 90$. The algebraic order of convergence is close to 4. This was expected as the temporal scheme is the RK4 method at order 4. Moreover, at this early stage of the simulation the error decreases with a power 4 law only when $C_o$ $\gtrsim 1.0$. The lowest errors are achieved at $C_o$ $\approx 0.75$. Below that $C_o$ number, a threshold is met: the error remains constant when the time step (and $C_o$ number) is further decreased; it is then controlled by the spatial discretization. 
It is also possible to note that the CFL number $C_o$ seems to be a relevant metric when testing the convergence of the method: the range of $C_o$ in which the results converge is the same across the 4 considered spatial discretizations.  

\begin{rmk}
    Significant differences in terms of $C_o$ with the work of \citet{HPC:hanssen2017free} have to be stressed. In their simulations the chosen numbers of points per wavelength were similar to the ones used here ($N_x\in[15,90]$), but the time step was constant and fixed at a small value of $\delta t/ T=1/N_t=1/250$. This value yields a $C_o$ between 0.06 and 0.36. This range of $C_o$ was shown to be out of the domain of convergence in time in our case. For the same $C_o$ (and apparently the same RK4 time scheme), the computation is indeed converged with respect to the time discretization and yields low error during the first periods, but instabilities then occur when the wave are freely evolving on a longer time scale. \citet{HPC:hanssen2017free} also encountered instabilities with this IB method. To counteract these instabilities, they used a 12\textsuperscript{th} order Savitzky-Golay filter in order to suppress, or at least, attenuate them. No filtering nor smoothing was used in our simulations. This may explain the differences of behavior with \citet{HPC:hanssen2017free} in terms of $C_o$ number.  Similarly, in \citet{hpc:zhu2017improved}, also with the RK4 time scheme, the time-step is chosen as $\delta t/T=1/200$ for a spatial discretization of $\delta x = h/10$. Converted to our numerical case, this would correspond to $N_x=100$, and so a $C_o$ fixed at $0.5$.
    During the investigations of \citet{HPC:ma2017local} on periodic wave propagation, their spatial discretizations ranged from $N_x=16$ to $N_x=128$. The equivalent $C_o$ number is thus comprised between $0.4$ and $3.2$. 
\end{rmk}

\begin{figure}[!htbp]
    \centering
    \includegraphics{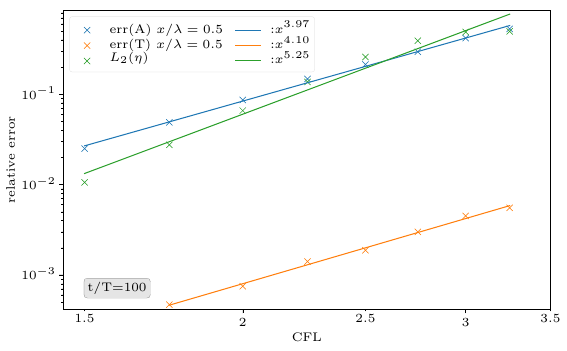}
    \caption{Convergence of the error on wave period and amplitude of the wave elevation at $x/\lambda=0.5$ for $N_x=30$. The $L_2(\eta)$ error -combination of both- is also added.}
    \label{fig:tempx32}
\end{figure}

\subparagraph{}
At long time $t/T=100$ (see \cref{fig:sub:convt100}), the error behaves differently. First, the error is approximately one order of magnitude higher compared to the time $t/T=1$, but the convergence rate is also slightly different, actually higher. 
As a matter of fact, at $t/T=100$, an order 4 of convergence is still found on the wave period and on the amplitude of the computed wave: \cref{fig:tempx32} 
shows the dependence in CFL of the error in %
 wave period and amplitude at the center of the domain $x/\lambda=0.5$ at $t/T=100$ for a fixed $N_x=30$ as a function of the CFL number $C_o$. The reference case used here is the one with $N_x=100$ at $C_o=0.05$ computed on one period. On a given case, the period is computed through the mean time separating two successive maximums, then a sliding Fast Fourier Transformation (FFT) is performed to obtained an accurate estimation of the amplitude of the free surface elevation.

However, the behavior of the $L_2(\eta)$ error results from a combined effect of both the error on the wave period and the error on the amplitude. The relative effect of those errors on the total error is analyzed in detail in appendix~\labelcref{apd:convergencelongt}, and a brief summary is given here.
Let $e$ be the relative error between two cosine functions. The first is the target function and the second one tends to the first one in amplitude as $\epsilon_a=f_A d^4$ and in period as $\epsilon_t=f_Td^4$. Here $d$ is a discretization variable -either $C_o$ or $1/N_x$ in our case-, which drives the convergence. $f_A$ and $f_T$ are constants with respect to $d$. %
At a whole number of periods -- \ie $t/T\in\mathbb{N}$ -- a Taylor expansion of $e$ when $d\to0$ can be performed:

\begin{equation}
    e  =  f_A d^4+  2\pi^2\dfrac{t^2}{T^2} f_T^2 d^8 +  (2\pi \dfrac{t}{T}f_T^3  - 2 f_A \pi^2\dfrac{t^2}{T^2} f_T^2 ) d^{12}  +O(d^{16})
    \label{eq:errordev}
\end{equation}
Note that $f_A$ depends on $t/T$ because the error on amplitude increases with time (in practice a linear dependence was observed at long time, \ie $f_A=\bar{f_A} t/T$). However $f_A$ and $f_T$ should not depend on the convergence parameter $d$.
The order 8 of convergence should disappear for small enough $d$ whatever $t/T$. In that case, the error on period is negligible compared to the error on amplitude: this results from the presence of the cosine, which elevates the error to the power 2.
However, if the error in amplitude $f_A$ increases in time slower than $t^2f_T^2$, there exists a time after which the error on the wave period will play an important role (order 8 will be predominant). This effect is thought to explain the seemingly high order of convergence of the $L_2(\eta)$ error in \cref{fig:sub:convt100}. \Cref{apd:convergencelongt} shows detailed comparisons at $t/T=100$ with values of $f_A$ and $f_T$ extracted from our results.

\subsection{Convergence with spatial discretization \label{ssec:standing_w_conv_x}}
\begin{figure}[htbp!]
	\centering
    \subfloat{\label{fig:sub:convx1}\includegraphics{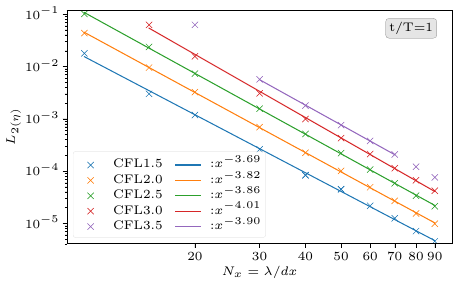}}
    \subfloat{\label{fig:sub:convx100}\includegraphics{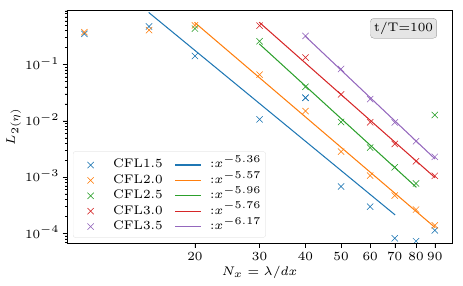}}
    \caption{Convergence of the $L_2$ error on $\eta$ in mesh refinement, with temporal discretization fixed. Solid lines correspond to the power regression of the error.}
    \label{fig:convergencext1t100}
\end{figure}

The convergence with spatial refinement  (\ie as a function of $N_x=\lambda/\delta x$) is analyzed in the same way and shown in \cref{fig:convergencext1t100}. The order of convergence in space is again 4.
Due to the choice of the set of HP including polynomials up to order 4, and the fact that the finite difference scheme of order 4 is used to compute the derivatives of free surface variables, this order 4 was the expected order of convergence.

At long time the convergence rate exhibits the same behavior as shown in the convergence with time resolution. The latter comments concerning the long time evolution of the error still holds (with, here, $d\equiv \delta x$), and is again thought to explain the increasing order of convergence of the total error $L_2(\eta)$ with time. 
Of course the values of the corresponding constants $f_A$ and $f_T$ are different.
Another effect also occurs: when the $C_o$ number increases, so does the order of convergence. This small yet clear effect at time $t/T=100$ is not completely understood. It would mean that the value of $f_T$ increases faster with the $C_o$ number than the value of $f_A$ does.

        \subsection{Effect of filtering the free surface variables} \label{ssec:filtering}
\begin{figure}[htbp!]
	\centering
    \includegraphics{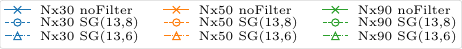}

    \subfloat{\label{fig:sub:convt1filter}\includegraphics{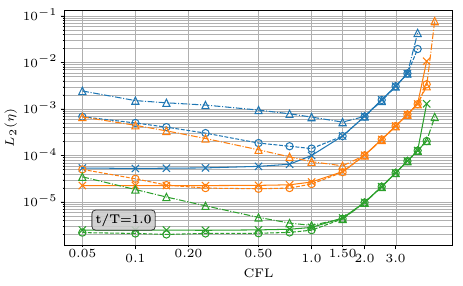}}
    \subfloat{\label{fig:sub:convt100filter}\includegraphics{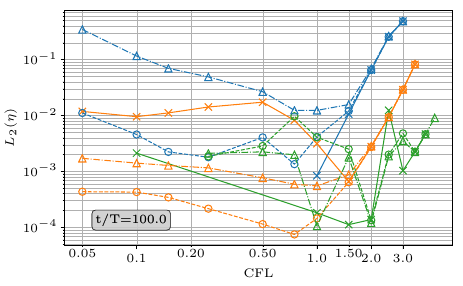}}
    \caption{Convergence of the $L_2$  error on $\eta$ with respect to the CFL number at two different physical times: $t/T=1$ (left panel) and $t/T=100$ (right panel), showing the effect of Savitzky-Golay filters SG(13,6) (triangles) and SG(13,8) (circles) in comparison to simulations run without any filter (crosses).}
	\label{fig:1comp2DErrtLog}
\end{figure}

As shown in the previous sub-section, stable computations are obtained for a range of spatial and temporal discretizations without any filtering. The phenomenon limiting this range is the so-called "sawtooth instability" developping on the free surface immersed boundary. In this sub-section, we evaluate the effect of applying a  centered  Savitzky-Golay filter \citep{savitzky_smoothing_1964} on the achievable range of discretization parameters. This filter is denoted SG($N$,$O$) where $N$ is the total number of points of the filter (\ie $(N-1)/2$ points are involved on each side of a considered node) and $O$ its order. An in-depth investigation on the effect of such filter is given in \citet{ducrozet_non-linear_2014}.

The filter is applied to the free surface elevation first. Afterwards, a correction step of the free surface potential was shown to be necessary to avoid the growth of instabilities: the free surface potential is enforced from an interpolation using the LE \cref{eq:mainwithcij} at the updated point elevation. %
The same filter is then applied to the newly obtained free surface velocity potential.
This procedure is applied once per time step, after the resolution of the main BVP. 

\Cref{fig:1comp2DErrtLog} shows the dependence on CFL of the free surface elevation error after 1 and 100  wave periods without filter (similar curves as in \cref{fig:convergencett1t100}) and with two SG filters with $N=13$ and orders $O=6$ and $8$ respectively. A great improvement on the width of the accessible parameter range is obtained, especially for the long simulation period $(t/T=100)$, for which almost all the simulations ran to their target duration. 

It can be noted that no significant gain in terms of accuracy can be obtained at short time $(t/T=1)$. Rather, the coarsest mesh resolutions exhibit a significant increase in total error. When increasing the spatial grid step $\delta x$ at fixed CFL value, two opposite effects are present: 1) the time step $\delta t$ increases, leading to a reduced number of filter applications for a given simulation duration, 2) however, each application of the SG filter increases in strength as shown by \citet{ducrozet_non-linear_2014}. Thus, as the total error increases as the mesh gets coarser, it is deduced that the second effect is dominant over the first one. This effect is more marked with SG(13,6) in comparison with SG(13,8). %
It is also possible to observe the former effect at fixed $N_x$ with decreasing CFL, as the total error increases. 

After a longer time evolution, benefits of the filter can be pointed out: lower CFL can be used without breaking the simulation. For some spatial discretizations, this leads to a lower minimal achievable error. For example, at $N_x=50$ the simulation at CFL $=0.75$ is stabilized, leading to the best achieved accuracy, slightly better than with $N_x=90$, CFL $=2$.

Overall, this type of filter, in particular used with SG(13,8) setting, appears to be able to contain the growth of instabilities and allows to run simulations with large intervals of discretization parameters. Though an increase in error is noted in some cases characterized by low CFL values at short times, the long term results are improved regarding both the number of stable simulations and the final error levels.  

Because good enough stability was denoted in a relatively large range of $N_x,CFL$ when the simulated time is contained ($\leq100T$), the use of a filter was not found to be necessary for the rest of the computations presented in this paper, unless stated otherwise.

\subsection{Highly nonlinear standing wave \label{ssec:highlynonlinearstanding}}
Though the standing wave case considered above has a rather high steepness $(H/\lambda=10\%)$, in order to approach the maximal achievable standing wave height, a test case inspired from \citet{ducrozet_non-linear_2014} was simulated. A nonlinear regular wave train with incident wave steepness $H_I/\lambda_I=7.5\%$ is generated at the left boundary of the domain using a stream-function solution of order 20. These waves are fully reflected by a vertical wall at the right boundary. The incident wavelength is set to $\lambda_I=1$ m for a constant water depth of $h/\lambda_I=1$. A SG$(13,8)$ filter was found to be necessary to simulate such level of nonlinearity.

\begin{figure}[htb!]
    \centering
    \includegraphics{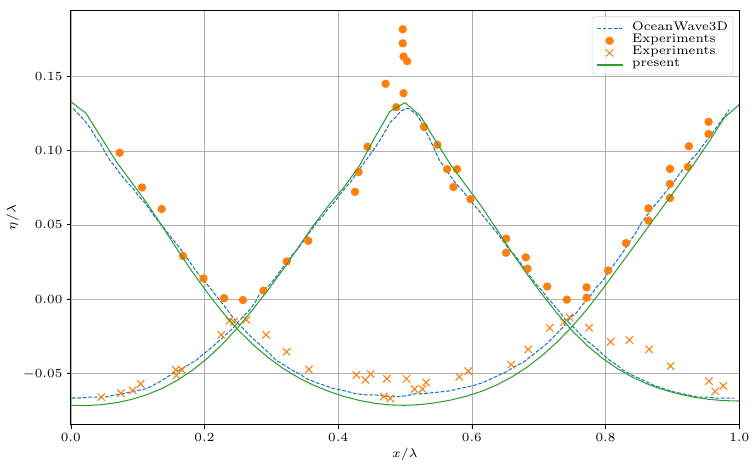}
    \caption{Free surface elevation profiles at time instants corresponding to minimum and maximun amplitude at an antinode position (\ie at $x/\lambda=0.5$) of the extreme standing wave generated by a full reflection on a vertical wall of an incident wave with $H_I/\lambda_I=7.5\%$. Comparisons are made with the experiments of \citet{taylor_experimental_1953} (with filled circles for data from the downward pointing probe and crosses for data from the upward pointing probe) and the numerical results of OceanWave3D taken from \citet{ducrozet_non-linear_2014}.}
    \label{fig:plotVsDucrozet}
\end{figure}

\Cref{fig:plotVsDucrozet} depicts the obtained wave profiles over one wavelength close to the reflecting wall at the peak and though time instants compared the experiments of \citet{taylor_experimental_1953} and OceanWave3D results presented by \citet{ducrozet_non-linear_2014}. A good agreement with both sources of data is obtained, with very limited differences with OceanWave3D. The obtained steepness, relative to the standing wave wavelength, is larger than twice the incident one, and is very close to the theoretical maximum of $H_S/\lambda_S\approx 0.20$.  

\subsection{Note on the computational efficiency \label{ssec:standing_comp_cost}}

\begin{figure}[htb!]
    \centering
    \includegraphics{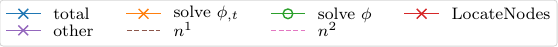}
    \subfloat[$N_x=60$, domain width varying]{\label{fig:sub:plotMultiWidthCosts}\includegraphics{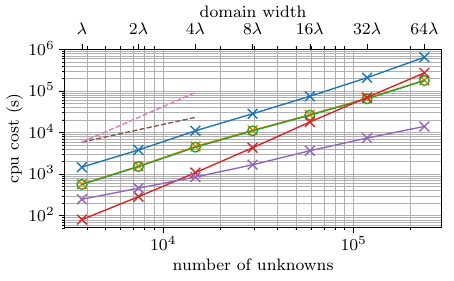}}
    \subfloat[$1\lambda$, spatial discretization varying]{\label{fig:sub:plotMultiNxCosts}\includegraphics{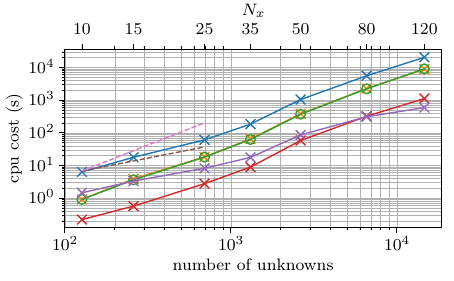}}
    \caption{
    Computational cost of 100 wave periods with different mesh widths (\labelcref{fig:sub:plotMultiWidthCosts}) with $N_x=60$, and different spatial discretizations $N_x$ for a domain width of one wavelength (\labelcref{fig:sub:plotMultiNxCosts}). The CFL number is here fixed at $2$. All computations are run in sequential mode on an Intel\textregistered{} Xeon\textregistered{} CPU E5-2650 v3  at 2.30 GHz
    }
    \label{fig:plotCosts}
\end{figure}

\Cref{fig:plotCosts} shows the computational burdens of the current implementation of the NWT on the standing wave case of \Cref{ssec:standing_w_intro}. \Cref{fig:sub:plotMultiNxCosts} represents some of the CPU costs encountered during the convergence study (CFL $=2.0$ on \cref{fig:convergencext1t100}) with a one-wavelength long domain and different values of $N_x$, while \cref{fig:sub:plotMultiWidthCosts} is representative of more classical applications, simulating a domain of variable length (from 1 to 64 wavelengths) and a fixed value $N_x=60$.

Note that many routines could still be optimized, especially the mesh related ones. For example, the current implementation of the routine that determines whether a node is outside or inside the water domain (``LocateNodes'' on the figure), is currently of complexity $O(N_{bb} N_s)$ where $N_{bb}$ is the number of nodes within the smallest box that contains all the free surface nodes and $N_s$ is the number of segments on the free surface. When only the domain width is varying, this reduces to a complexity of $O(N^2)$ where $N$ is the total number of nodes, which is confirmed on \cref{fig:sub:plotMultiWidthCosts}.

While not needed in the standing wave case because no loads are to be computed, the problem on $\phi_t$ is resolved anyway to give a better grasp of what the computational costs would be in a real application case. The problems on $\phi$ and $\phi_t$ yield the exact same matrices, and their resolutions are thus of same cost. However, one could take advantage of the first resolution. This constitutes a second example of possible optimization of the current code. 

When the spatial discretization is reduced, the CFL being fixed implies that the time step decreases. Thus, the cost of computing a given number of periods is increased by a factor $N$: the time spent on each routine is increased by both the number of points but also the number of times it is repeated.

From \cref{fig:sub:plotMultiWidthCosts}, a rough estimate of the CPU time required for the simulation of $t/T$ periods on a real application case (of domain width $8\lambda$) and water depth $h$, assuming that the problem is of complexity $O(N)$, can be given as:
\begin{equation}
    c_{CPU}= \dfrac{t}{100T} \dfrac{h}{\lambda}c_{s,t/T=100,8\lambda}
    \label{eq:roughcpu}
\end{equation}
where $c_{s,t/T=100,8\lambda} = 2\cdot 10^4~\text{s}=5~\text{h}~33~\text{min}$, \ie the CPU cost for the simulation of 100 periods of the standing wave case on a domain of width $8\lambda$ with a water depth of $h=\lambda$. Of course, this expression constitutes a crude approximation and different routines, corresponding to the methods presented later in \cref{sec:theory_fixedBody}, might increase the CPU cost of the overall simulation, even more so if they are not optimized.

\subsection{Summary of numerical convergence study \label{ssec:standing_w_sum}}
After a comprehensive numerical study of the HPC method on challenging nonlinear standing wave cases, the efficiency and accuracy of the IBM modeling of the free surface applied on a fixed underlying spatial mesh was demonstrated. Optimal ranges of spatial and temporal discretization parameters were determined:
\begin{itemize}
    \item the spatial discretization $\delta x$ should be chosen to have $N_x=\lambda / \delta x$ between $40$ and $90$ nodes per wavelength, which is a reasonable range of values for practical applications.
    \item the temporal discretization $\delta t$ should be best selected to have a CFL number $1.5 \le C_o \le 3.5$, meaning that the number of time steps per period $N_t=T/\delta t$, also given by $N_x/C_o$, is then comprised between $N_x/3.5$ and $N_x/1.5$. This is highly beneficial as it authorizes rather large time-steps for practical applications.
    \item Moreover, the application of a Savitzky-Golay filter broadens this stable range for long simulation duration, especially for low CFL numbers, allowing to reach smaller total error at the cost of an increased CPU time.  
    \item the present implementation of the HPC shows an algebraic convergence rate with the spatial resolution of order greater than 4.
    \item it also shows an algebraic convergence rate with the temporal resolution of order comprised between 4 and 5 (for long time simulations).
    \item extrapolating the obtained computational burdens predicts a contained cost of the numerical model.
\end{itemize}
Finally, the computational burden is shown to remain limited, even for relatively long simulation duration. This is true even when a domain representative of an engineering application case is selected.

\section{Introduction of a fixed body in the NWT \label{sec:theory_fixedBody}}
\subsection{A double mesh strategy to adapt the resolution in the vicinity of a body \label{sec:sub:twoMeshIntro}}
After having validated the method with nonlinear waves in the previous section, and particularly the immersed free-surface strategy, the next objective is to include a body in the fluid domain, either fully submerged or floating. Obviously, a desirable solution would retain cells of square shape and constant geometry as much as possible, even in the case of a moving body (not treated here however).

Once again, different strategies are possible. \citet{HPC:hanssen2015harmonic} first introduced an IB method for bodies in waves in the HPC framework. \citet{HPC:ma2017local} compared this method with an immersed overlapping grid fitted to the boundaries (corresponding to the body in our case). Recently, this technique was successfully applied to a 3D problem by \citet{liang_liquid_2020}: the liquid sloshing problem in an upright circular tank. Accurate results were obtained when compared to weakly nonlinear and linear model theories. This newly introduced grid will often be referred to as the "fitted mesh" for simplicity.
In the current study, the latter strategy is chosen. Two main reasons led to this choice: first, an oscillatory behavior was exhibited by \citet[Figs. 24,25]{HPC:ma2017local} when studying the spatial convergence with an IBM. This oscillatory behavior is also present when the body is moving, with a large magnitude. This is mainly due to an incremental change in the chosen ghost nodes which can turn to be favorable at some time steps (\ie for certain grid configurations) and unfavorable at some other ones. Moreover, this oscillatory behavior does not come with a reduction of the error, both for the fixed and oscillating body.

The second reason is that adding a new fitted mesh allows to decouple the discretization of the wave propagation part %
(usually defined with respect to the wavelength, \eg approximately $N_x \in [40-90]$ as shown in the previous section) from the discretization appropriate for the resolution of the potential close to the body (usually defined with respect to the body characteristic dimension, denoted $D$). Thus, by using two different grids, a suitable discretization for both the wavelength and the computation of the loads would be possible. For instance, including a small body relative to the incoming wavelength would be challenging with the IBM: too many nodes would be required so as to correctly solve the BVP in the vicinity of the body whereas less nodes would be needed further away. From a quantitative point of view, the case inspired from \citet{chaplin1984nonlinear} and treated in \cref{ssec:chaplin} hereafter involves an important ratio $\lambda/D\approx 15$. Thus, if the far-field discretization is set as $N_x=90$ to correctly capture the propagation of the waves, then $D$ is discretized with only $6$ nodes. Such high values of the $\lambda/D$ ratio, often encountered in practical engineering applications, are more easily taken into account with a second grid fitting the body than with a choice of higher order cell or with a local refinement of the grid. 
Note that the solution combining both a secondary grid and a solid immersed boundary has been tested by \citet{HPC:hanssen2019non,hanssen_potential_2021} with promising results. On this subject \citet[\S~4.2.2]{HPC:ma2017local} applied this combination to model the free surface and obtained a important reduction of the resulting error. 

\subsection{The two way-communication inside the fluid domain \label{sec:sub:twoMeshComm}}
\begin{figure}[htbp!]
	\begin{center}
    \includegraphics{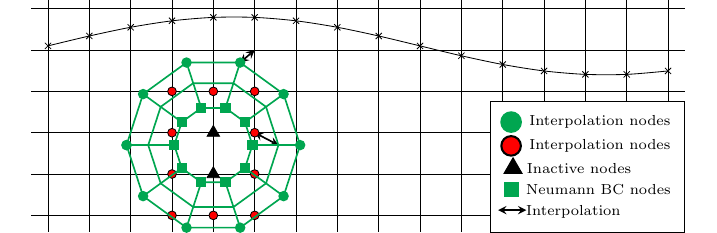}
    \caption{Schematic representation of the immersed free surface and body below the free surface. A circle means an "interpolation" node (described in the text) while the colors are used to identify particular nodes on the two grids, as indicated in the legend.}
	\label{sch:twoMeshesCom}
	\end{center}
\end{figure}

Thus, a boundary fitted grid (BFG) is added locally around the body, overlapping the background grid (BGG). These grids and the points associated to the method are represented on \cref{sch:twoMeshesCom}. The Laplace problem is solved on both these grids simultaneously, \ie both domains are solved in the same global matrix problem. The global matrix size is increased by the number of nodes of the BFG and decreased by the number of nodes inactivated in the BGG. One can note that the global matrix is thus almost defined by block, each corresponding to a grid.

Thus, the boundary nodes of the new BFG also need a dedicated equation in order for the system to be closed.
For a node laying on the body boundary, a simple Neumann boundary condition is set and enforced in the global matrix, as described in \cref{sec:sub:dirandneucond}.
The nodes indicated by circle markers on \cref{sch:twoMeshesCom} serve as communication nodes between the two grids and are denoted ``interpolation nodes''. For example, for a node 
$P_f$ located on the outer contour of the BFG (green circle markers on \cref{sch:twoMeshesCom}),  the imposed equation in the global matrix is the interpolation equation from the closest macro-cell in the BGG (\cref{eq:mainwithcij}). 
On \cref{sch:twoMeshesCom}, a double arrow gives a representative example of the link between $P_f$ and the center of the closest macro-cell in the BGG. 
This ensures -- in a implicit manner -- that the potential at the location $\tI{x}_{P_f}$ is the same in both meshes:

\begin{equation}
    \phi^{(f)}_{P_f}  =\phi^{(bg)}(\tI{x}_{P_f})
    \label{eq:communitationInterpolationFromBg} 
\end{equation}
where $\phi^{(f)}_{P_f}$ is the potential of the particular node $P_f$ (directly an unknown of our system of equations), and $\phi^{(bg)}(\tI{x}_{P_f})$ represents the value of the interpolation equation from the background potential field at the given coordinate $\tI{x}_{P_f}$. Further developing \cref{eq:communitationInterpolationFromBg} and using the closest macro-cell local expression (\cref{eq:mainwithcij}) yields an implicit interpolation equation:

\begin{equation}
    \phi^{(f)}_{P_f}  =\left. \sum_{i=1}^8 \left(\sum_{j=1}^8 C_{ji}^{-1} f_j(\tI{\bar{x}}_{P_f})\right)  \phi_{P_i}\right|_{(bgc)}
\end{equation}
Where the notation $\left.\right|_{(bgc)}$ emphasizes that the local expression is applied on the closest background macro-cell. $\phi_{P_i}$ are the potential of the bounding nodes of this macro-cell. %
Note that the two problems associated with the different nodes are almost independent. The global matrice is thus almost defined by block, except at those interpolation nodes%
: the potential of a node $P_f$ in the fitted grid is implicitly linked with the potentials of the neighboring background nodes.

Points belonging to the BGG situated inside the body are inactivated (black triangles on \cref{sch:twoMeshesCom}). 
    Thus, equations are needed for the points of the BGG surrounding those inactive points (red circle markers on \cref{sch:twoMeshesCom}).
The same method is used here: the interpolation equation~(\cref{eq:mainwithcij}) is enforced such that the interpolation of the fitted grid potential matches the node potential. In other words, the interpolation is effective from the BFG to the BGG, using the LE of the closest cell of the fitted mesh. Thus, denoting this point $P_{bg}$ and its coordinates $\tI{x}_{bg}$:

\begin{equation}
    \phi^{(bg)}_{P_{bg}}=\phi^{(f)}(\tI{x}_{P_{bg}})
    \label{eq:communitationInterpolationFromFitted}
\end{equation}
Here again, this relation is represented for one particular node on \cref{sch:twoMeshesCom} by a double arrow.

So, by considering the various type of nodes discussed above, the proposed method ensures a consistent implicit two-way communication between the two meshes of interest, as the BVP problems (on $\phi$ and $\phi_t$) are solved on both grids simultaneously, \ie through the inversion of a unique global linear problem.
\subsection{Free surface piercing body \label{sec:sub:Fscom}}
\begin{figure}[htbp!]
	\begin{center}
        \includegraphics{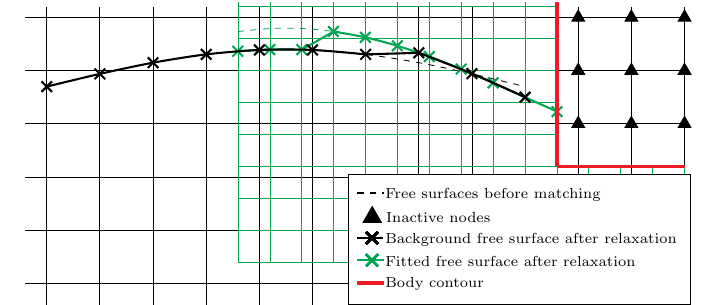}
        \caption{Schematic representation of the background free surface and the fitted free surface. A matching is enforced between the two free surfaces.
    The difference between the two free surfaces is here exaggerated for clarity.}
	\label{sch:twoMeshesFs}
	\end{center}
\end{figure}

At this stage, a two way communication is ensured between the fitted grid (BFG) and the background grid (BGG). %
The problem is closed in the sense that every node involved in the global matrix has its own dedicated equation. %
Still, a difficulty arises when the fitted mesh pierces the free surface. %
Indeed, it is not possible to interpolate outer points of the fitted mesh where no solution is computed (above the free surface): %
this issue is solved by introducing a new free surface, evolving in the BFG, which %
allows to solve and advance the free surface locally at the scale of the body.  In this work, having a dedicated discretization in the vicinity of the body was proven to be necessary, when for example the impact of the incident wave on the front of the body resulted in the generation of waves of short wavelength and large steepness. %
Note that using a single free surface with different set of markers is also possible \citep[see \eg][]{HPC:hanssen2019non,hanssen_potential_2021}. %
The free surface evolving in the background grid is truncated such that no marker is defined inside the body (\ie markers are only present in the fluid domain) as can be seen in \cref{sch:twoMeshesFs}.

For simplification purposes, the new free surface evolving in the BFG will be called "fitted free surface" even though this free surface also uses the IBM described in \cref{sec:sub:ifs}. 
Thus, we obtain two free surfaces, with different resolutions in space, following their respective grid discretizations, that overlap each other in the vicinity of the body. 
To ensure the communication between both free surface curves, the outer nodes positions and values of variables $\phi$ and $\phi_t$ of one free surface are interpolated and enforced through a 1D B-spline interpolation from the other free surface. Referring to the schematic representation in \cref{sch:twoMeshesFs}, the position and values of the outer right node of the background free surface is enforced so as to match the fitted free surface. Reciprocally, the position and values of the outer left node of the fitted free surface is enforced so as to match the background free surface.   

However, if this enforcement affects only one marker at the extremity, instabilities may occur. For example a stencil of two points on each side is the minimal length to maintain a 4\textsuperscript{th} order of spatial convergence with a 1D centered finite difference scheme.
To prevent this from having an important impact, relaxation zones are set to incrementally match the free surfaces at their extremities. Relaxation formulas and weights are thus needed for every marker:

\begin{equation}
    \gamma_e = (1-\alpha) \gamma_i + \alpha \gamma_t \label{eq:relaxationzoneslocal}
\end{equation}
 where $\gamma_e$ represents the value to enforce, $\gamma_i$ the initial marker value, $\gamma_t$ the target value (interpolated value from the other free surface) and $\alpha$ an arbitrary weight function of the marker position that evolves between $0$ and $1$. Note that for this application, $\gamma$ stands for either $\phi$, $\phi_t$ or $\eta$.   
Many different functions for $\alpha$ were tested and implemented without significant impact. In practice free surfaces are completely matched over a given length (\ie $\alpha=1$ if the marker distance to the free surface extremity is lower than a certain threshold, $\alpha=0$ otherwise). The \cref{sch:twoMeshesFs} emphasizes the effect of such relaxation functions: %
it shows the free surfaces before (dotted lines) and after the matching (solid lines) using this method. Other methods exist for this type of matching. For example, in the context of a domain decomposition approach, a buffer zone on which a common free surface is computed and enforced in both domains is presented in \citet{kim_simple_2010} and \citet{lu_overlapping_2017}. This latter approach, involving only one buffer zone, may be regarded as a particular case of the method presented here.
    Another hybrid coupling between the OceanWave3D and OpenFOAM codes that uses relaxation zones to transfer information from one model to the other was recently investigated by \citet{kemper_development_2019}. For a more detailed review of coupling methods, the reader is referred to \citet{di_paolo_wave_2021}, and in particular their Table 1.

Note that if the body is not at the extremity of the computational domain, a second fitted free surface is needed on the other side of the body. This will be used in \cref{ssec:barge_ECM}. 

From a numerical point of view, the fitted mesh is considered as an unstructured grid. At a price of additional coding efforts and an increase of CPU time when identifying points (as well as an increase in the memory usage), a gain is made on the simplicity of inclusion of complex bodies of arbitrary shape. However, as already stated, the implemented HPC method requires square cells to be most effective \citep{HPC:ma2017local}. 
Taking advantage of the fact that the BGG will remain a structured mono block, it would thus be possible to modify the methods on this grid to reduce both the RAM requirement and the necessary CPU time associated with identification of node types and interpolations between the resolutions of the BVP themselves.

\subsection{Singular node treatment (sharp corners) \label{ssec:singularNodes}}
\begin{figure}[htb!]
    \centering
    \includegraphics{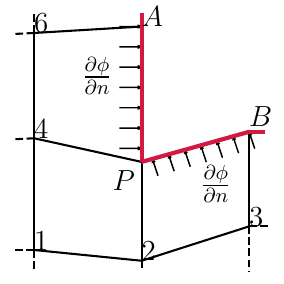}
    \caption{Illustrative sketch of a sharp body or domain corner.}
    \label{fig:sharpcorners}
\end{figure}

\Cref{fig:sharpcorners} depicts a sharp corner situation. In this situation the choice of the Neumann direction to impose a wall condition at the sharp corner node (denoted $P$ on the figure) is not obvious. However, in the classical HPC approach, only one equation can be set for that node. 

To overcome this issue, a first approach was suggested by \citet{liang2015application}. They applied a domain decomposition method in the vicinity of the corner apex: a local corner-flow solution was then coupled with the HPC problem.

Another method, suggested by \citet{hpc:zhu2017improved}, consists in imposing a null total flux through the line APB. In the current study, an extension of this method is implemented to tackle external corners such as body corners. In this framework, the obtained and enforced equation links the unknowns appearing in two different macro-cells. This equation is $F_{APB}=0$ where, considering straight segments, the total flux can be expressed as:

\begin{align}
    F_{APB}
     & =
    l_{PA} \int_{PA} \left(\nabla \phi(\tI{ x}(s)) \cdot \tI{n}\right) ds 
    +
    l_{PB} \int_{PB} \left(\nabla \phi(\tI{ x}(s)) \cdot \tI{n}\right) ds  \\
     & =
    \left.
        l_{AP} \sum_{i=1}^{8}  \left(\sum_{j=1}^{8} C^{-1}_{ji} \tI{I}_j (AP) \cdot \tI{n}(AP) \right) \phi_i 
    \right|_{c_{AP}}
    +
    \left.
        l_{PB} \sum_{i=1}^{8}  \left(\sum_{j=1}^{8} C^{-1}_{ji} \tI{I}_j (PB) \cdot \tI{n}(PB) \right) \phi_i 
    \right|_{c_{PB}} 
    \label{eq:mainfluxderivated}
\end{align}
where $\tI{I}_j(AP)=\int_{AP} \nabla f_j(\tI{ \bar{x}}(s)) ds$, $s$ is the scaled abscissa varying in $[0,1]$ within the given segment, $\tI{n}(AP)$, $\tI{n}(PB)$ are the normal vectors to the segments $AP$ and $PB$ respectively. The length of a segment is denoted with $l$. A vertical bar is added to underline that each segment considers a different macro-cell for $\phi_i$, $\tI{I}_j$, $C_{ij}$, $\tI{n}$, and for the computation of the local coordinate $\tI{\bar{x}}(s)$. 
Note that the vector integrals $\tI{I}_j$ can be computed analytically, knowing the analytical expressions of the harmonic polynomials. 

\begin{figure}[htb!]
    \centering
    \includegraphics[width=0.4\textwidth]{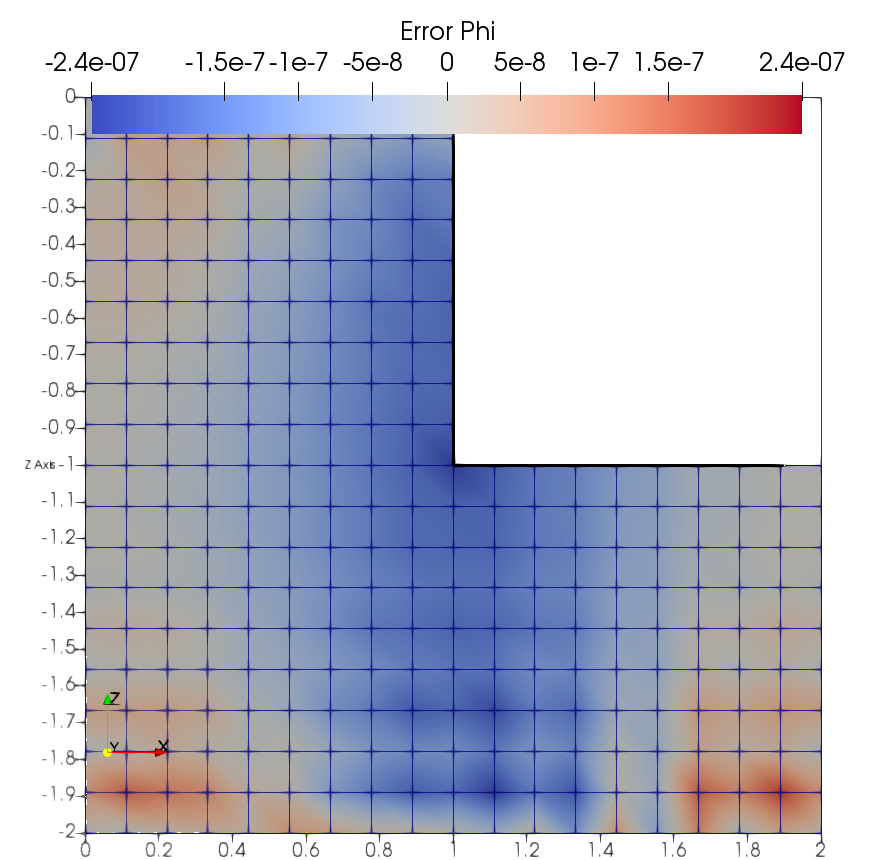}
    \caption{
        Sharp external body corner. Local error on the potential $(\phi - \phi_{th})/\max(\phi_{th})$. $\phi_{th}=\cosh(k(z+h/2))\cos(kx)$ where $k=2\pi/L$. $L =2$~m and $h = 2$~m are respectively the total width and height of the domain. This analytical function is imposed as Neumann condition on the black lines and as Dirichlet condition at the other boundary nodes. The sharp corner method is used on the center point of coordinates $(L/2,-h/2)=(1,-1)$.
    }
    \label{fig:sharpcornersRes}
\end{figure}

The implementation of this method is assessed on a simple case shown in \cref{fig:sharpcornersRes} for which the boundary conditions are set according to an arbitrary analytical function, given in the legend of the figure. This figure shows the normalized error on the potential over the domain after the solution of the Laplace BVP. It can be observed that the error in the vicinity of the corner is of similar magnitude as other nodes in the domain, which validates the implementation. 
In practice this method wad selected for mostly three reasons: i) it removes the need of an arbitrary choice of normal vector at the corner node, ii) it results in a better total mass conservation, and iii) it increases the numerical stability on real wave-body interaction cases (not shown here).

\section{Validation of wave-body interaction against two flume experiments \label{sec:valid}}
In order to validate the methods presented above, two experimental test cases were selected. The first one is chosen to verify the boundary-fitted overlapping grid method with a fully submerged horizontal cylinder of circular cross-section. The second one is a free surface piercing barge of a rectangular cross-section. 
\subsection{Fixed horizontal submerged cylinder \label{ssec:chaplin}} 
\citet{chaplin1984nonlinear} studied a fixed horizontal cylinder, with a low submergence below the SWL, in regular waves of period $T=1$~s. Accurate experimental results about the nonlinearities of wave loads on the cylinder were obtained and are often used to validate NWTs \citep[\eg][]{guerber_modelisation_2011}. The total water depth is $d=0.85$~m which, together with the period, imposes a wavelength of approximately $\lambda=1.56$~m (slightly varying with the incident wave height).
The cylinder of diameter $D=0.102$~m is immersed with its center located at $z_c=-D$ below the SWL.

This problem is numerically difficult to solve for volume field methods as the cylinder is close to the free surface, such that the fluid domain right above the cylinder is reduced to a small water gap of height $D/2 \approx \lambda /30$ (when the water is at rest). This water gap needs to be meshed and resolved with the HPC method. Thus, a spatial discretization of $\lambda/\delta x\in [40,90]$ would yield a discretization of this gap with only $\sim 2-3$ nodes.

\begin{figure}[htbp!]
	\begin{center}
        \includegraphics[width=0.95\textwidth]{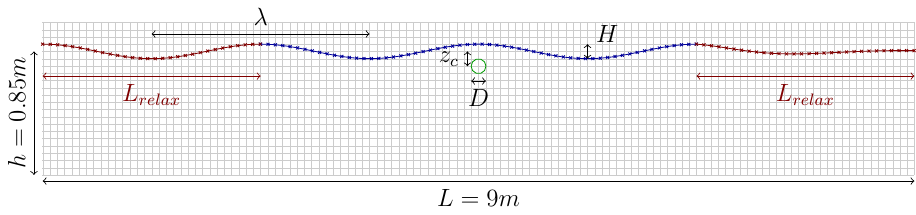}
        \caption{Schematic representation of the numerical set-up inspired from \citet{chaplin1984nonlinear}. The local refined mesh fitted to the cylinder (not represented in this figure) is shown in \cref{fig:plotMeshesChaplin}.}
	\label{sch:cylinder}
	\end{center}
\end{figure}

The nonlinear regular incident waves are generated using the so-called stream function theory \citep{fenton1988numerical}.
To avoid reflection on both the wave maker side and the outlet Neumann wall and to impose the target incident wave field, relaxation zones are introduced to enforce the requested values over a distance chosen as $L_{relax}=\lambda$ through a commonly used exponential weighting function \citep[\eg][]{jacobsenFuhrmanFredsoe2012}. Note that other techniques of waves generation and absorption are possible. For example \citet{clamond_efficient_2005} introduced a damping term in the Bernoulli equation in order to modify the DFSBC: the wave elevation is smoothly driven to the SWL. \Cref{sch:cylinder} shows at scale the computational domain including the two relaxation zones.
	
\begin{figure}[htbp!]
    \centering
    \includegraphics{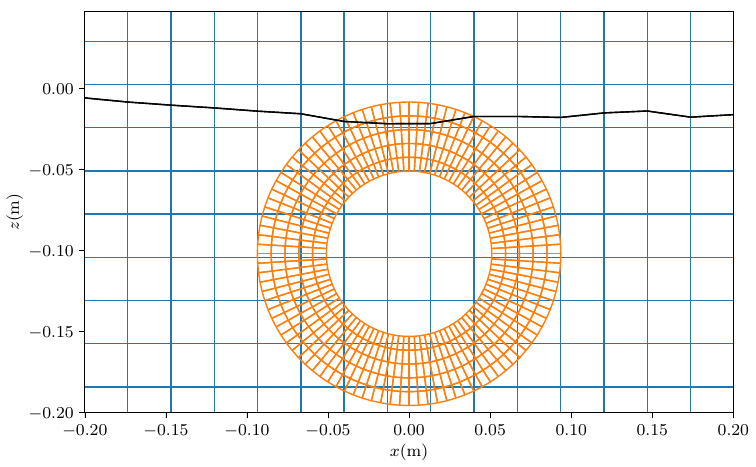}
    \caption{Example of mesh grids and free surface at the passage of a wave trough with $H/\lambda=2.3\%$. The SWL is at $z=0$~m.}
    \label{fig:plotMeshesChaplin}
\end{figure}

\begin{figure}[!htbp]
\centering
    \includegraphics{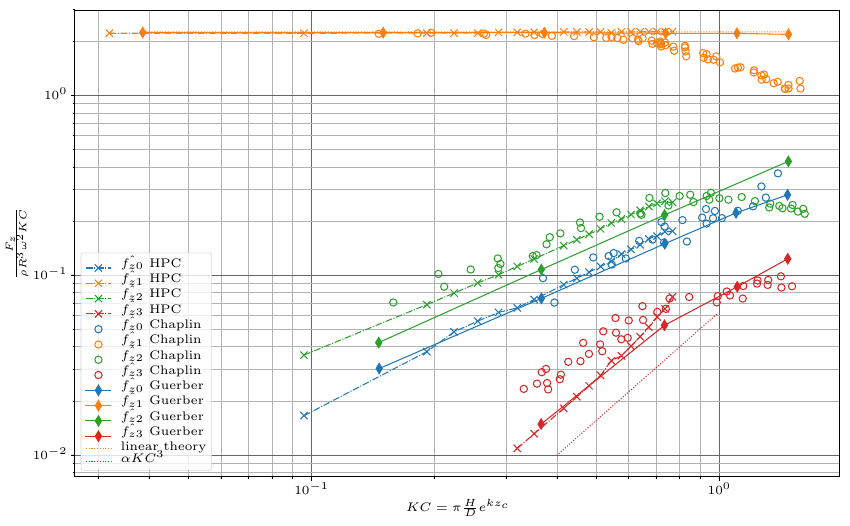}
    \caption{Amplitudes of the various harmonic components of the vertical load on the horizontal circular cylinder. Current results (crosses) compared with numerical simulations from~\citet{guerber_modelisation_2011} (lines with diamonds) and experiments from~\citet{chaplin1984nonlinear} (empty circles). The amplitude of the first order harmonic based on linear prediction is added as well as a third order model line, to compare with the evolution of the amplitude of the third order harmonic.}
    \label{fig:Error2DCyl}
\end{figure}

We focus our attention to the vertical force exerted on the cylinder once the periodic wave motion is established in the NWT. A Fourier analysis is applied to the computed time series of vertical force. The normalized amplitudes of the harmonics of the vertical force and the mean vertical force (drift force) are plotted in \cref{fig:Error2DCyl} as a function of the Keulegan-Carpenter number ($KC$), defined as: %

\begin{equation}
    KC= \pi\frac{H}{D}\exp(k z_c)
    \label{eq:KC}
\end{equation}
On this figure, the results from the present NWT are compared with the experimental values from \citet{chaplin1984nonlinear} and the numerical results from \citet{guerber_modelisation_2011}. The linear theory results from \citet{ogilvie_1963} are also added for comparison of the amplitude of the first harmonic.

All harmonics amplitudes up to third order are in relative good agreement with the BEM simulations from \citet{guerber_modelisation_2011}, although some discrepancies can be denoted. %
The mean value and first order harmonic are difficult to distinguish between the two sets of numerical results, whereas the amplitude of the second order harmonic is closer from the experimental data with the current HPC method. For the third order harmonic, which is of very small relative amplitude, it is difficult to identify the most accurate method.

Moreover, the behavior of the different harmonics amplitudes seems to agree with the expected theoretical results: the amplitude of the first harmonic increases as $KC^1$ (\ie $\hat{f_{z1}}/KC$ is constant), the drift force and second order harmonic amplitude both increase as $KC^2$, and the third order harmonic amplitude increases as $KC^3$. Regarding the latter, the HPC method reproduces more closely that order 3 in $KC$ in comparison with both \citet{guerber_modelisation_2011} and the experimental results. 
Of course, limitations in the comparison with the experiments can be observed, in particular for larger wave heights. This is mainly due to the viscous effects (not considered here, nor in the simulations of \citet{guerber_modelisation_2011}). When viscous simulations are done, the experimental variation can be retrieved \citep[\eg][]{tavassoli2001interactions}.

However, a difficulty arises with the HPC method when $KC$ increases (\ie for larger wave heights). As the Laplace equation is solved in the different cells inside the fluid volume, there should always be at least a fluid point in the volume above the cylinder top. In waves of large amplitude, the cylinder top is very close to the free surface, in particular when a wave trough passes over the cylinder. In this situation, it is not possible to keep the number of points per wave length $N_x$ in the previously selected range $[40,90]$ and keep square cells. Saw-tooth instabilities appear as $KC$ exceeds $0.80$ approximately. No filter was used in this case, as the main objective is to emphasis the limits of the method itself. Above a wave steepness of $H/\lambda = 2.6\%$, (\ie $KC=0.86$), no computation could remain stable after $\sim 2-3$ wave periods.

As a conclusion, even if a limitation in wave height is met, those results, correct up to third order, give us confidence on a case which is particularly challenging for volume field methods. 

\subsection{Rectangular barge, experiments and numerical comparison \label{ssec:barge_ECM}}

In order to evaluate the NWT and its limitations with a (fixed) free surface piercing body, dedicated experiments were conducted in a wave flume at Centrale Marseille with a barge of rectangular cross-section. 
\subsubsection{Experimental set-up}
\begin{figure}[!htbp]
\centering
\includegraphics[width=0.60\textwidth]{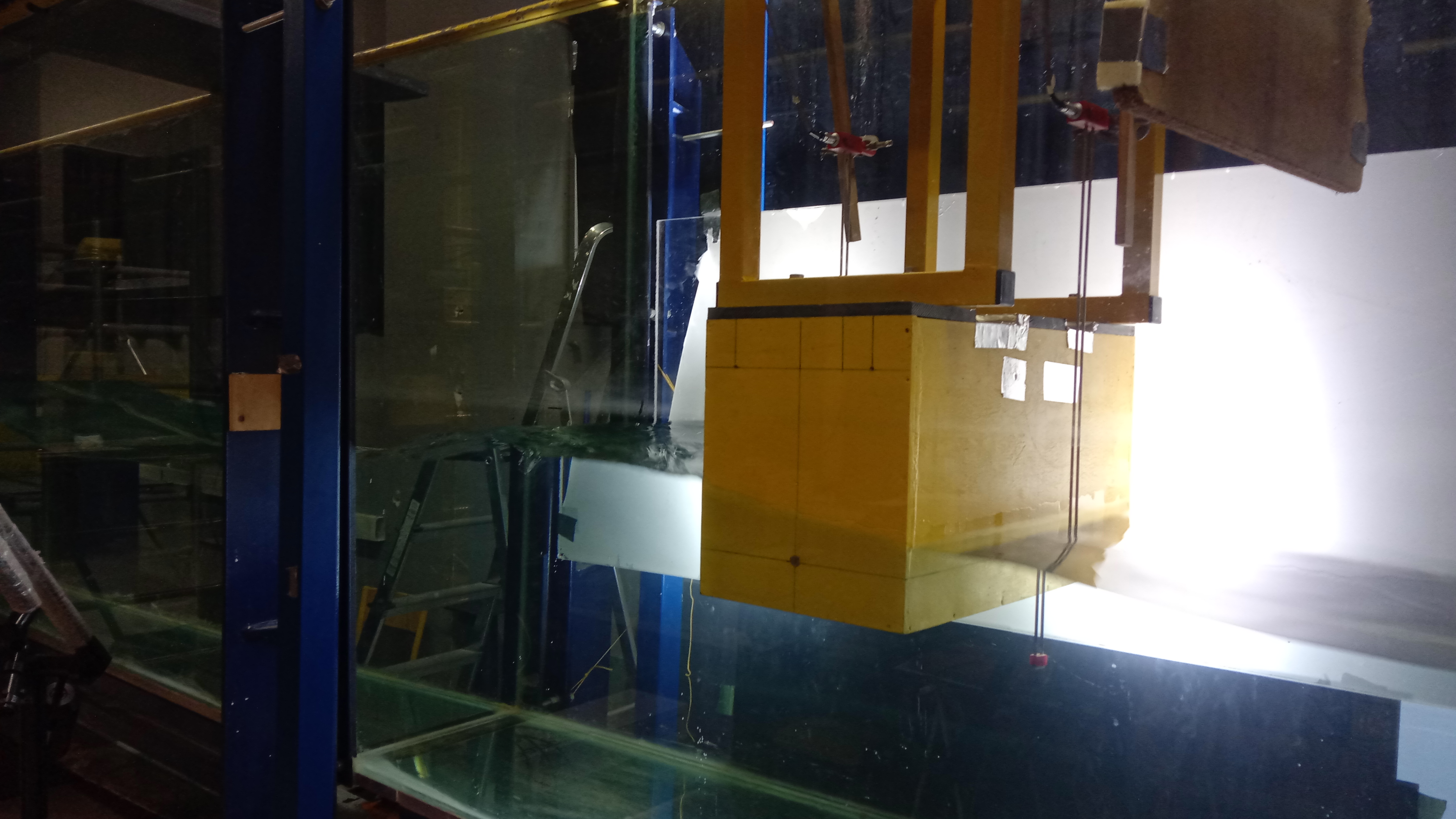}
\caption{Photograph of the experimental set-up of the rectangular barge in the wave flume (the wave maker is on the left side).}
    \label{fig:Experimentbarge}
\end{figure}

The flume is $17$~m long and $0.65$~m wide. The water depth was set to $d=0.509$~m. The body is a rectangular barge of draft $0.10$~m for a length of $0.30$~m (in the longitudinal direction of the flume) mounted on a 6-axis load cell measuring device. The width of the barge spans the width of the flume minus a small water gap of about 2~mm on both sides between the barge and the flume side-walls (\cref{fig:Experimentbarge}). A perforated metallic beach is placed at the end of the flume to dissipate the energy of the transmitted waves, and so to avoid reflection. The waves are generated with a flap type wave maker.
The body is placed such that its front face is located at $x_b=11.52$~m from the wave maker. 13 wave gauges are installed all along the wave flume. Unfortunately, the wave gauge dedicated to measure the run-up on the front face of the barge was found \emph{a posteriori} to be defective. For some cases, a video of the experiment was recorded in the vicinity of the body, allowing to extract free surface profile and run-up at the front and rear faces of the barge.

\begin{figure}[!htbp]
\centering
\includegraphics{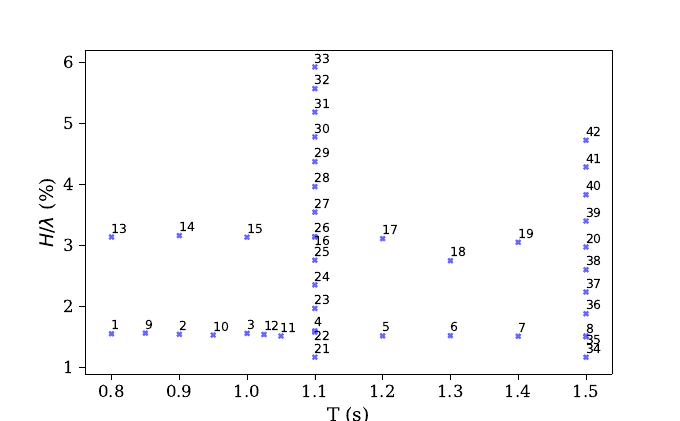}
\caption{Overview of the conducted experiments in the $(T, H/\lambda)$ plane.}
    \label{fig:tabexperiment}
\end{figure}

Approximately 40 cases were tested in regular wave conditions, with varying wave period and height, as illustrated in \cref{fig:tabexperiment}. A focus is made here on two periods: $T=1.1$~s and $T=1.5$~s . For the latter one, however, high wave steepness yielded high wave heights, resulting in dewetting and breaking, with a lot of turbulent effects, recirculations, and air entrainment. Thus, for this period, few relevant comparisons can be made with the potential model (which neglects all those effects).

\subsubsection{Numerical set-up}
The computational domain is automatically generated and meshed by the NWT depending on the case: its length is chosen as $\sim 12\lambda$ and relaxation zones are set in the inlet and outlet, over a length of $2\lambda$. A second mesh is fitted to the body in order to ensure a precise computation of the flow dynamics in its vicinity. The mesh fitting the body is of breadth $0.30$~m on each side (\ie a total extent of $0.90$~m). Both meshes communicate through previously described interpolation boundaries. Their free surface curves communicate through relaxation zones of length $0.14$~m and constant function of unitary weight. In order to interpolate a free surface from the other one, a 1D B-spline interpolation is used. Different discretizations of the fitted mesh were tested, without significant difference.

\subsubsection{Numerical results and comparisons with experiments} \label{ssec:resultsbarge}
For the wave period $T=1.1$~s, the steepest cases (31-33) were found to numerically break down. This is due to an large run-down which leads to a dewetting at the bottom left corner of the barge. During the experiments, recirculations and turbulent effects were clearly visible for those steepest cases. 

\begin{figure}[!htbp]
\centering
\includegraphics{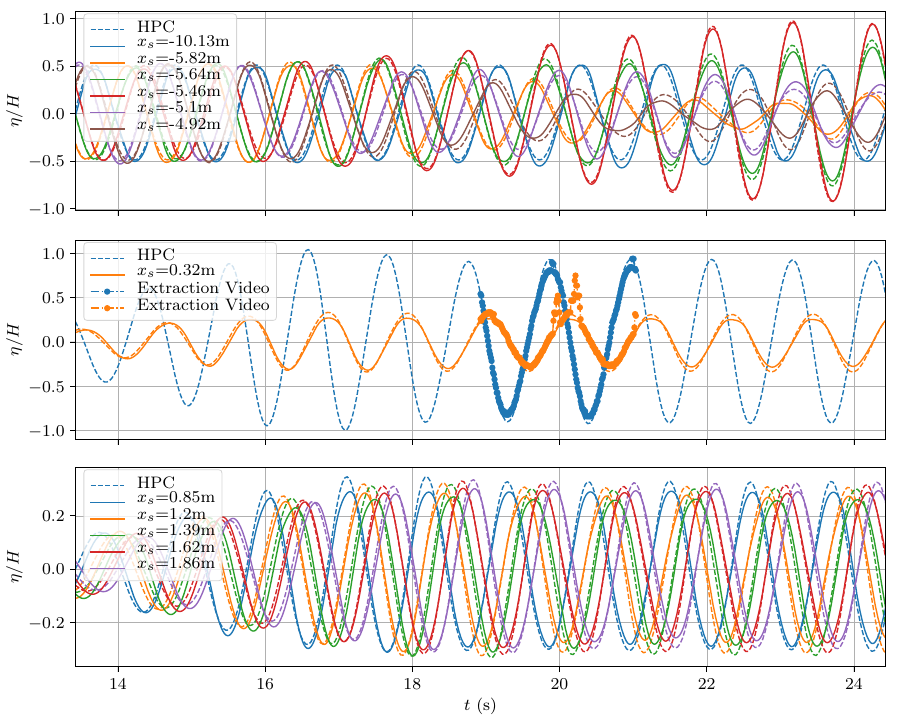}
\caption{Comparisons of the free surface elevations recorded at different positions: HPC (dashed lines), experimental wave gauges (solid lines) and extracted elevation from the experimental video (circle markers). Wave gauges are divided in three groups (corresponding to subfigures): incident waves upstream (subfig 1), front and rear run-up (subfig 2), transmitted waves (downstream waves, subfig 3). Case 21: $T=1.1$~s, $H/\lambda=1.2\%$.}
    \label{fig:SondesCase21}
\end{figure}
\begin{figure}[!htbp]
\centering
\includegraphics{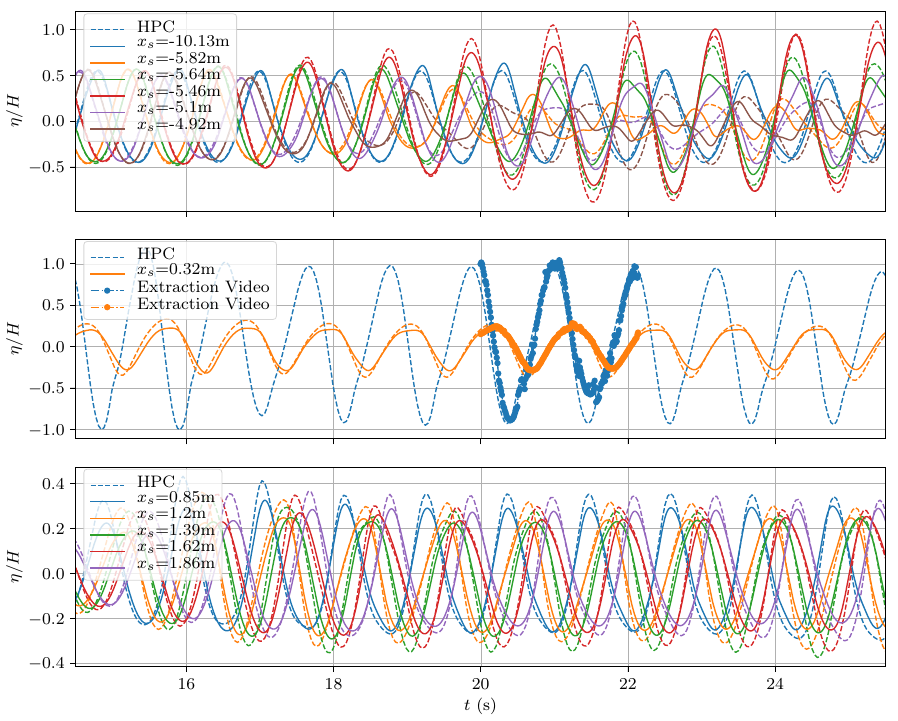}
\caption{Comparisons of the free surface elevations recorded at different positions: HPC (dashed lines), experimental wave gauges (solid lines) and extracted elevation from the experimental video (circle markers). Wave gauges are divided in three groups (corresponding to subfigures): Incident waves upstream (subfig 1), Front and rear run-up (subfig 2), transmitted waves (downstream waves, subfig 3). Case 30: $T=1.1$~s, $H/\lambda=4.8\%$. }
    \label{fig:SondesCase30}
\end{figure}

On \cref{fig:SondesCase21,fig:SondesCase30}, wave gauge measurements are compared to the free surface elevation time series from the HPC method for cases 21 and 30 respectively. Note the time synchronisation between measurements and simulations was done on the first gauge only, and the determined phase shift was then applied to all the remaining gauges.
This means that relative phases of the wave elevations in the numerical simulations are represented adequately. More generally, a very good agreement is found on both wave amplitudes and phases, at the various locations along the wave flume. However, some discrepancies of the run-up at the front and rear faces of the barge as well as concerning the downstream wave elevations can be observed. Those discrepancies become more marked when the wave steepness increases. It can be observed that the HPC model tends to overestimate larger run-up events, and transmitted wave heights.  
The underlying causes of those discrepancies are thought to be related to the mathematical model itself. It is well known that non-dissipative models cannot correctly describe the flow in the vicinity of sharp angles even for linear incident waves. Thus, the present potential model is not perfectly appropriate to model the behavior of this type of flow. In practice, this effect is often counteracted by introducing a numerical lid in the vicinity of the body to numerically dissipate some energy (so trying to mimic dissipation due to viscosity). With such a method, the length and strength of the lid need to be tuned to match the expected result. This option was not tested here.

    Reflection ($C_r$) and transmission ($C_t$) coefficients (in terms of wave height ratios) have been calculated on a 2-period time window ($t\in[23~\text{s},25.2~\text{s}]$) for the case 21 (\cref{fig:SondesCase21}) using the nonlinear method of \citet{andersen_estimation_2017} with 5 wave probes on each side of the body. We obtain $Cr=81\%$ from the experiments and $C_r=79\%$ from the numerical predictions.
    In the same manner the transmission coefficients are in good agreement: $C_t=60\%$ and $62\%$ in the experiment and the numerical simulation respectively.

Obviously, the observed discrepancies in terms of run-up are expected to impact the loads exerted on the barge, mainly through the difference of run-up on the front and rear faces, impacting for example the elevation at which the pressure is set to the atmospheric pressure (dynamic boundary condition at the free surface).

\begin{figure}[!htbp]
\centering
\includegraphics{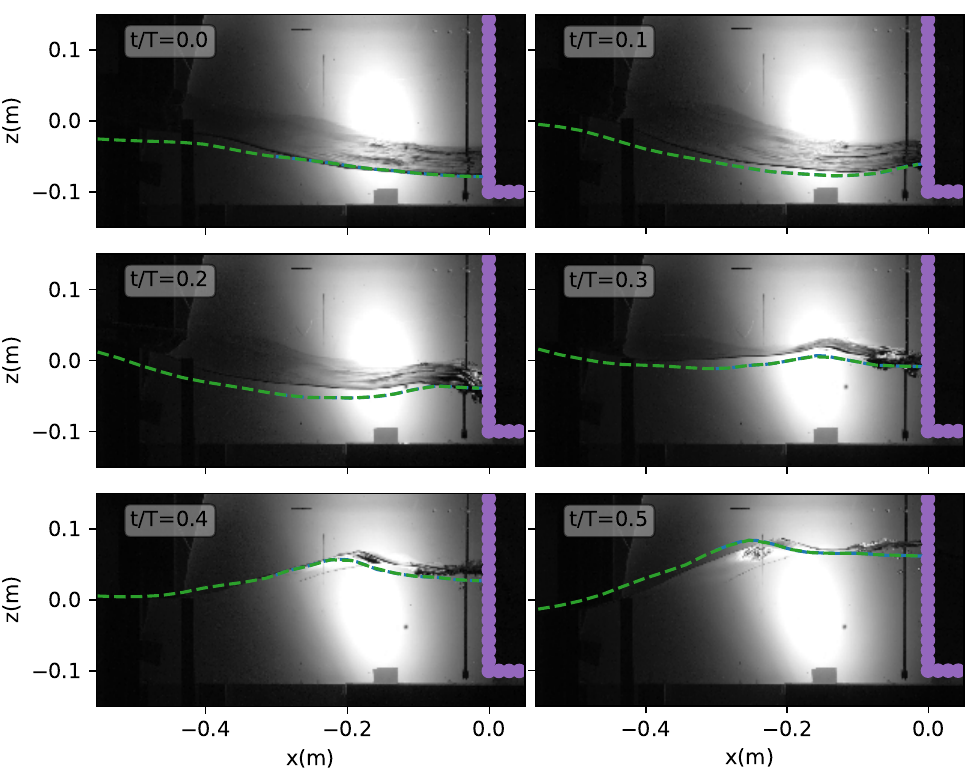}
\caption{Case 30: $T=1.1$~s, $H/\lambda=4.8\%$. Comparison of the computed free surface elevation (green dashed lines) in front of the rectangular barge with the experiment at six different time instants. Incident waves come from the left.}
    \label{fig:compareFreeSurface30}
\end{figure}

\Cref{fig:compareFreeSurface30} shows at different time instants the free surface elevation in front of the body obtained from the HPC computation superimposed on snapshots from the experiments. On these pictures, we clearly denote aspects which cannot be taken into account in the potential model: during the rise of the water level, air entrainment and wave breaking take place, leading to important complex turbulent effects. At this stage, it is expected that the viscous effects play an important role. Although those dissipative effects are neglected in the HPC model, a relative fair agreement can be seen on the figure. The run-up is approximately correctly captured, as well as the reflected wave emerging during the elevation of the run-up. %
Moreover, the numerically computed reflected wave seems to be slightly faster than the experimental one. The difference can again be attributed to viscous effects, delaying the apparition of the reflected wave.

\begin{figure}[!htbp]
\centering
\subfloat[Case 21]{\label{fig:sub:f21}\includegraphics{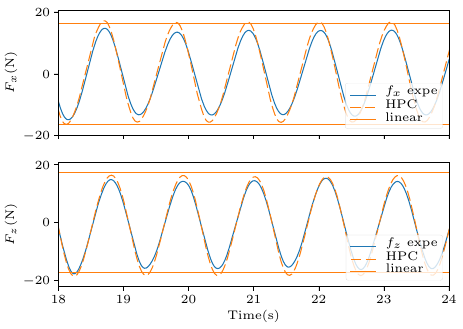}}
\subfloat[Case 30]{\label{fig:sub:f30}\includegraphics{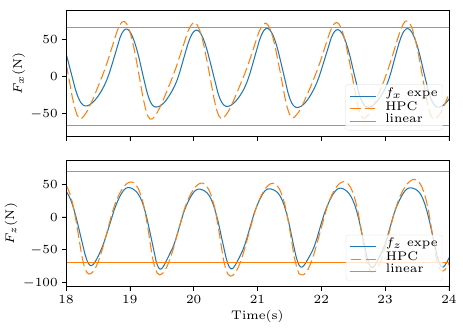}}
\caption{Cases 21 (left) and 30 (right): $T=1.1s$, $H/\lambda=1.2\%$ and $H/\lambda=4.8\%$. Comparison of time series of the hydrodynamic loads on the barge: experiments, linear results and HPC simulations. }
    \label{fig:temporalLoads21}
\end{figure}

Time series of horizontal ($F_x$) and vertical ($F_z$) force components on the body are depicted on \cref{fig:temporalLoads21} for the cases 21 and 30, respectively the most linear and nonlinear cases with $T=1.1$~s. The horizontal orange lines (symmetric with respect to the zero-force line) represent the amplitude of linear predictions of these loads.
When the incoming waves are close to be linear (case 21), a good agreement is found between the linear and HPC models: the amplitudes are almost equal though the mean value computed with the HPC model slightly moves the vertical load extrema from the symmetric horizontal lines of the linear model result. 
Note that the effective (local) steepness, which determines the degree of nonlinearity, close to the body is approximately twice the incident steepness $H/\lambda$: only a small part of the energy is transmitted and thus the reflected wave adds to the incident wave locally. For a quantitative analysis, the amplitudes of incident and transmitted wave elevations can be compared in \cref{fig:SondesCase21}.

When compared with experimental data, the computed loads are of approximately $15\%$ higher magnitude.
Note that, from an engineering point of view, the model is conservative in this case, and as expected, the differences in terms of run-up on the front and rear sides of the body lead to a slight overprediction of both the horizontal and vertical loads. This overprediction is seen for all cases, and consistently with approximately the same relative value.
For steep incoming waves, the discrepancies between the linear and nonlinear models increase, as one could expect. In particular, the mean value (drift force) is no longer null. For case 30, one can note that, although the loads are overpredicted in magnitude, the shape of the time series of the loads (and thus the magnitude of the higher order harmonics of the loads) seems to be in good agreement with the experiment. 

\begin{figure}[!htbp]
\centering
\includegraphics{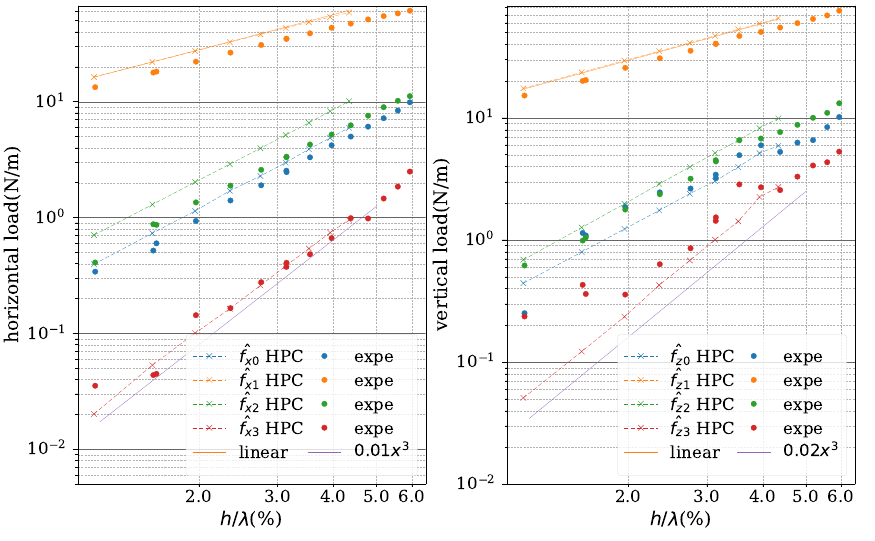}
\caption{$T=1.1$~s, Amplitudes of harmonics of the vertical and horizontal loads on the barge. HPC models (dashed lines), experimental results (dots) and linear transfer functions (solid orange line). A power function of order 3 is added to compare the increase rate (solid purple line).}
    \label{fig:harmonicsT11}
\end{figure}
In order to compare the computed loads with the experimental ones more precisely, a Fourier decomposition of the time series is performed. The Fourier coefficients of this decomposition are shown for all the cases performed at $T=1.1$~s on \cref{fig:harmonicsT11}. 
As expected the amplitudes of the first harmonic exhibit an over-estimation compared to the experimental data, whereas the mean values ($\hat{f}_{xz}^0)$ are in correct agreement. 
However, the behavior of all the harmonics with the wave height is found to be exactly the same as measured during the experiments (experimental and computed lines are parallel).  
For the vertical loads, all harmonics are roughly consistent with the measurements.
An important difference can be noticed for the third order harmonic of the vertical load for small steepness. First, it can be denoted that this harmonic is about 2 orders of magnitude lower than the first order harmonic, and even 3 orders magnitude lower than the hydrostatic force. Thus, an accurate computation as well as an accurate measurement of this small contribution is difficult to achieve. Moreover, the variations of the potential NWT values seem to be more closely fitted with a model of order 3 in wave steepness. This observation, together with the fact that numerical and experimental third order harmonics match for higher wave steepness, can be regarded as an indication of model's capability to resolve high-order nonlinear effects. Two reasons might explain these discrepancies: 1) either the measurements, the numerical method, or the post-processing method are not accurate enough to resolve or compute such small contributions, or 2) a physical effect bringing energy to the third harmonic plays a role not properly taken into account by the model.

For the horizontal loads, more discrepancies are visible. First order and drift force show an overestimation of about 15\%. The amplitude of the third order harmonic seems to be well captured, but the amplitude of the second order one exhibits a larger over-estimation by about 70\%. This effect is still under investigation, but the impact on the free surface of the viscous effects seems to be significant in this case. Note that the hydrostatic contribution originating from instantaneous wet free surface is one of the main contributions to the second order component. In the present set-up, this term will play a role only on the horizontal load as the wet free surface changes only on the vertical walls of the barge. Thus, an important difference between the experiments and the numerical model is expected on this particular second order harmonic of the horizontal load: it was shown previously that our computations do not capture exactly the effective run-up elevations, probably due to the viscous effects not taken into account in the HPC model.

\section{Summary of conclusions and outlook} \label{sec:conclusions}
In this work, an implementation of the Harmonic Polynomial Cell (HPC) method for solving the fully nonlinear potential flow problem in combination with an Immersed Boundary Method (IBM) to track the free surface was achieved and validated. The newly developed Numerical Wave Tank (NWT) allows to simulate a wide range of situations in ocean and coastal engineering applications, involving nonlinear waves and wave-body interaction. In this version, the bodies can be immersed or surface piercing, but have to be fixed.  The numerical techniques have been described, in particular the implications of choosing to work with non-deforming grids which has lead to the development of an accurate and robust IBM variant. 

After investigating the convergence properties of the numerical methods with respect to a refinement in space and time on a standing wave of steepness $H/\lambda = 10\%$, freely evolving over up to 100 wave periods, it was shown that the NWT exhibits a high level of accuracy and other interesting features. First, a large range of time steps are shown to lead to stable computations over the target duration, although for small CFL $C_o$ numbers the simulations tend to become unstable when no filter is applied. 
Most importantly, the recommended range of CFL was shown to be $C_o \in [1.5, 3.5]$, which is of great benefit as these large values permit to use large time steps, and so to reduce the computational burden. Regarding space discretization, the number of nodes per wavelength should lie in the range $[40,90]$. On this case, results are found to be accurate and converging with an order of convergence comprised between 4 and 5, when refining either the spatial or temporal discretization. An original theoretical investigation of the evolution of the numerical error as a function of the simulation duration was also proposed to explain the observed trends.

In addition, the application of a mild Savitzky-Golay filter (typically SG(13,8)) was shown to extend significantly the broadness of the stability domain. More specifically, smaller CFL numbers could be used, leading to a reduction of the minimal achievable error. 

Then, a variation of the standing wave case was set up to approach the maximum achievable wave height. Here, we simulated the reflection on a vertical wall of a nonlinear incident wave of steepness $H_I/\lambda_I =7.5\%$, leading to a standing wave of steepness reaching $H/\lambda \approx 20\%$, \ie close to the theoretical limit. The agreement of the present results with both experimental measurements and numerical results from another nonlinear NWT demonstrates the capabilities of the method. 

A second Boundary Value Problem is introduced and solved on the time derivative of the potential in order to obtain an accurate estimation of the pressure in the fluid domain through the Bernoulli equation. 
This method ensures a precise computation of loads on bodies, which will further be needed to compute the movements of freely moving bodies.


An multi overlapping-mesh method was also developed as a way to compute the flow accurately in the vicinity of an object subjected to incoming waves, using a finer local body-fitted grid. In case the body pierces the free surface, a second free surface is added close to the body and evolves in the body fitted mesh. Specific strategies to couple the two (here fixed) grids and associated (time varying) free surface curves were proposed. The proposed double mesh technique ensures an accurate computation of complex flow patterns arising due to the presence of the body. %
In addition, the method introduced by \citet{hpc:zhu2017improved} to treat the singular node problem was extended to internal body corners and was shown to yield  small volume errors on a simple dedicated BVP.

Coupled with relaxation zones to both generate and absorb waves at extremities of the NWT, the method is shown to be accurate for several 2D cases. In particular, the double mesh approach give accurate results on the case of a fully submerged horizontal cylinder located very close to the free surface \citep[flume experiments by][]{chaplin1984nonlinear}. We have shown that the mean (drift) vertical force on the cylinder and the amplitudes of the first three harmonics of the vertical force are properly estimated for a range of Keulegan-Carpenter number. The nonlinear capabilities of the model are thus confirmed. 

As a final test, the NWT is used to simulate dedicated experiments performed in the wave flume at Centrale Marseille with a surface piercing rectangular barge. As the present model does not consider viscous effects, this case proved to be really challenging, in particular in high wave conditions and due to the presence of sharp corners at the bottom of the barge. In these conditions, indeed, dissipative terms cannot be neglected if one wants an accurate prediction of the physical values.
Nevertheless, the NWT showed very good results for cases with low steepness incident waves (loads on the barge, run-up on the barge's vertical sides, transmitted and reflected waves). As the wave steepness increases, the simulations still reproduce the time evolution of free surface elevation, loads and run-up signals, but the differences in amplitude become more marked.  In these conditions, errors versus the experiments can be up to 15\% on the amplitude of the first harmonic of the loads. Higher harmonics tend also to be overestimated compared to the experimental results. We however would like to insist on the fact that some of the cases simulated here were extreme in the sense that waves started to break on the barge and we almost reached the dewetting of the front face of the barge (as shown in the pictures of the experiment in \cref{fig:compareFreeSurface30}). This being considered, the ability of the NWT to run on such cases and to deliver a correct (though slightly overestimated) order of magnitude of loads and run-up elevations is regarded as a valuable outcome of this study.

Future developments of this work will encompass four main aspects, namely: i) extension to 3D in order to simulate more realistic cases, ii) simulation of moving bodies in waves, iii) implementation a fully Lagrangian tracking method for the free surface nodes, and iv) coupling this potential NWT with local viscous models in the vicinity of bodies. This last development is meant to improve the simulation of cases where viscous effects are significant, as shown here for the rectangular barge.

\section*{Acknowledgments}
The participation and help provided by Dr. Olivier Kimmoun during the experiments conducted in the wave flume at Centrale Marseille and presented in \cref{ssec:barge_ECM} are gratefully acknowledged. The authors also express their gratitude to Pr. Bernard Molin for fruitful discussions during this work, and to the three anonymous reviewers whose comments helped improving earlier versions of this article.

\paragraph{Funding}
This work was supported by the Ecole Normale Sup\'erieure (ENS) de Cachan (France) in the form of a \emph{Ph.D.} grant attributed to Fabien Robaux. This research did not receive any other specific grant from funding agencies in the public, commercial, or not-for-profit sectors.
\pagebreak

\appendix
\section{Convergence of a perturbed cosine in both period and amplitude at long time}\label{apd:convergencelongt}

In order to analyze the rate of convergence of our implementation of the HPC method on the case of the nonlinear standing wave presented in \cref{sec:standing_w}, we consider that the theoretical (reference) solution for the free surface elevation at the center of the numerical wave tank, denoted $f_{th}(t)$, is a cosine function of period $T_{th}$ and amplitude $a_{th}$:

\begin{equation}
    f_{th}(t)=a_{th} \cos \left( 2\pi\dfrac{t}{T_{th}} \right)
    \label{eq:basecos}
\end{equation}
Assume then that the HPC solution for the free surface elevation at the same location can also be expressed  as a cosine function:

\begin{align}
    f(t) = a \cos \left(2\pi\dfrac{t}{T} \right)
    \label{eq:perturcos} 
\end{align}
with amplitude amplitude $a$ and period $T$, being close to $a_{th}$ and $T_{th}$ respectively.

Our numerical convergence tests have shown (see \cref{fig:tempx32}) that the errors on amplitude and period of the surface elevation at the center of the tank decrease as $C_o^4$. Lets generalize this with a more convenient parameter $d$ and thus assume that the error both in amplitude and period decreases as $d^4$. In this case $d\equiv C_o$ but the exact same reasoning can be applied with $d\equiv \delta x$. With this notation we have:

\begin{align}
a = & a_{th}(1-e_A)  = a_{th} ( 1 - f_A d^4 ) \label{eq:errorAmpl}\\
T = & T_{th}(1-e_T)  = T_{th} ( 1 - f_T d^4 )  \label{eq:errorPeriod}
\end{align}

Note that for every representation in this article, we set $f_A=-5.5\cdot10^{-3}$
and $f_T=-5.4 \cdot 10^{-5}$.
Those values are extracted from the convergence depicted \cref{fig:tempx32} with the aim of fitting the convergence of our HPC model on the standing wave for $d\equiv C_o$ and $t/T=100$.

\begin{figure}[htbp!]
    \centering
    \includegraphics{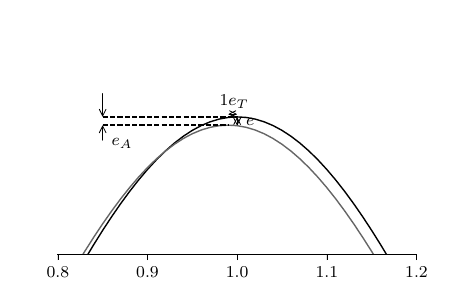}
    \includegraphics{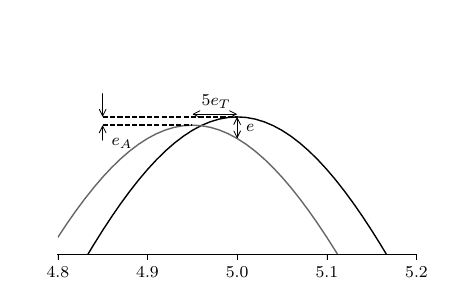}
    \caption{Sketch of the period, amplitude and total error of a perturbed cosine (grey) at two different whole periods. The assumption is made that $e_A$ and $e_T$ do not depend on time. }
    \label{fig:CosineAndPertubedCosineT1}
\end{figure}

Lets define $e$ the total relative error at a given time $t$:

\begin{align}
    e(t) & = \dfrac{f_{th}(t)-f(t)}{a_{th}} \label{eq:errortot} 
\end{align}
This error is the combined effect of both the error on amplitude and the error on period.
A representation of the impact of $e_A$ and $e_T$ and their combined impact - with the assumption that they are constant in time - are shown in sketch \cref{fig:CosineAndPertubedCosineT1}.

Developing this expression with our cosine models~(\cref{eq:basecos,eq:perturcos}), and injecting the modeled expression of the error~(\cref{eq:errorAmpl,eq:errorPeriod}) yields an expression of the total error as a function of $t$ and $d$.

\begin{align}
e(t) & =\cos(2\pi\dfrac{t}{T_{th}})-\dfrac{a}{a_{th}} \cos(2\pi\dfrac{t}{T})  \\
    e(t) & = (\cos(2\pi\dfrac{t}{T_{th}}) - \cos(2\pi\dfrac{t}{T}) ) + f_A d^4 \cos(2\pi\dfrac{t}{T}) 
\end{align}
This error is computed for different times $\dfrac{t}{T_{th}}$ and shown on \cref{fig:figtot} as a function of $d$. In the range $d\in[1.5,2.5]$ a power interpolation is computed and added.

\subsection{Taylor expansion} \label{apd:ssec:taylorexpansion}
In this part, a Taylor expansion of the latter expression of the total error is done in the vicinity of a whole number of period $t/T_{th}$:

\begin{align}
    e(t/T_{th}\in \mathbb{N})& = (1. - \cos(2\pi\dfrac{t}{T}))  + f_A d^4 \cos(2\pi\dfrac{t}{T}) 
    \label{eq:e1}
\end{align}

Let's first expand $t/T$ using the modeled behavior of T when $f_Td^4\to0$ the given in the expression~(\cref{eq:errorPeriod}):

\begin{align}
    2\pi\dfrac{t}{T} & = 2\pi \dfrac{t}{T_{th}} \dfrac{1}{1-f_Td^4} \\
    2\pi\dfrac{t}{T} & = 2\pi \dfrac{t}{T_{th}} ( 1 + f_Td^4 + f_T^2d^8+ O(d^{16})) \\
    2\pi\dfrac{t}{T} & = 2\pi \dfrac{t}{T_{th}}  + X  
\end{align}
where $X=2\pi \dfrac{t}{T_{th}} ( f_Td^4 + f_T^2d^8+ O(d^{16}))$. X also tends to 0, so it is possible to expand the cosine in the vicinity of its maximum (whole number of periods): 

\begin{align}
    e(t/T_{th}\in \mathbb{N}) & = (1 - [ 1-X^2/2+O(X^4)  ] ) + f_A d^4 [ 1-X^2/2+O(X^4)  ]   \\
    e(t/T_{th}\in \mathbb{N}) & = (2\pi^2\dfrac{t^2}{T_{th}^2} f_T^2 d^8+   2\pi \dfrac{t}{T_{th}}f_T^3d^{12}  +O(d^{16})  ] ) + f_A d^4 [ 1-2\pi^2\dfrac{t^2}{T_{th}^2} f_T^2 d^8 +O(d^{12})  ]  
\end{align}
At the end, we obtain the Taylor expansion of $e$ at order 12:

\begin{equation}
    e(t/T_{th}\in \mathbb{N})  =  f_A d^4+  2\pi^2\dfrac{t^2}{T_{th}^2} f_T^2 d^8 +  (2\pi \dfrac{t}{T_{th}}f_T^3  - 2 f_A \pi^2\dfrac{t^2}{T_{th}^2} f_T^2 ) d^{12}  +O(d^{16})
\end{equation}

\subsection{Representation and interpretation}
\begin{figure}[htbp!]
    \centering
    \includegraphics{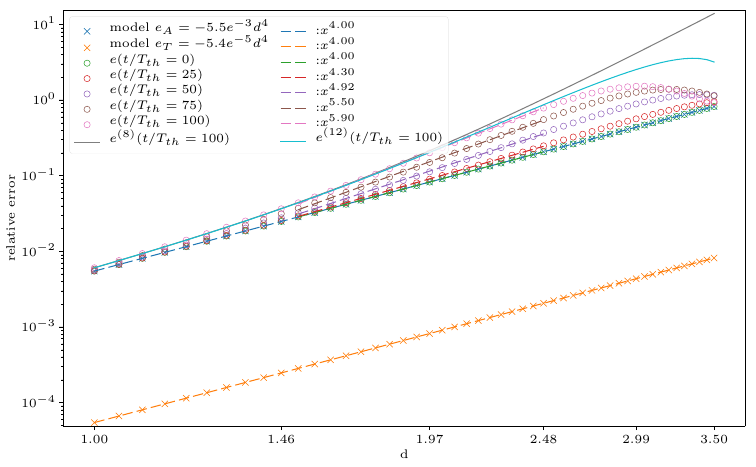}
    \caption{Relative error of models in period and amplitude ($e_T$ and $e_A$ respectively). Total relative error $e$  after $\dfrac{t}{T_{th}}$ periods between two cosine (\cref{eq:errortot}) as a function of $d$ which parameterize the order 4 convergence between their amplitudes and period. $e^{(n)}$ is the Taylor expansion of e at order $n$}
    \label{fig:figtot}
\end{figure}

\Cref{eq:errordev} is computed at $o(8)$ and $o(12)$ and represented as a function of $d$ on \cref{fig:figtot} (respectively labelled $e^{(8)}$ and $e^{(12)}$) at $\dfrac{t}{T_{th}}=100$. The parameters are kept fixed to $f_A=5.5\cdot 10^{-3}$ and $f_T=5.4\cdot 10^{-5}$ so as to fit the obtained results depicted on \cref{fig:tempx32}.
Thus, we model the case where $N_x/\lambda=30$ and the error is computed as the relative difference of the free surface elevation at the center point ($x/\lambda=0.5$) with $d\equiv C_o$.  Moreover, expression~(\cref{eq:errortot}) of $e$ is also shown at different times and for different convergence parameter in the range $[1, 3.5]$. 

For most lines a power regression is computed and added in the figure (dashed lines).
For the sake of simplicity, the values of $f_A$ and $f_T$ are supposed to be independent of $t/T_{th}$ and set from the observed convergence at $t/T_{th}$. In reality, this assumption is not verified. 
 
Note that the order of convergence is 4 when $\dfrac{t}{T_{th}}=0$ (\ie $t=0$). For $t\ne0$ the asymptotic error when $d$ converges to 0 is equal to the amplitude error. Mathematically, this is due to the square elevation when expanding the cosine. Thus, on \cref{fig:figtot}, the $d^4$ order model is not shown as it corresponds to the amplitude error.

In order to compare the different regimes of convergence, the ratio between the 8\textsuperscript{th} order and 4\textsuperscript{th} order is computed

\begin{equation}
    r_{8,4}=\dfrac{2\pi^2\dfrac{t^2}{T_{th}^2} f_T^2}{f_A}d^4
    \label{eq:ratio}
\end{equation}

If $f_A$ increases with time slower than $t^2f_T^2$, then the order 8 will have a growing importance over time ($r_{8,4}$ increase). As $f_A$, and $f_T$ where set to model the magnitude of the error after $t/T_{th}=100$ and $d\equiv C_o$, the ratio presented above is computed at this time step for different parameter $d$: 
8\textsuperscript{th} order and 4\textsuperscript{th} (\ie  $r_{8,4}=1$) are of the same importance when $\dfrac{t}{T_{th}} = 100$ as soon as $d=1.75$. For $d$ as small as $1.00$, the 8\textsuperscript{th} order already represents $r_{8,4}=10\%$ of the error. Thus, it is both mathematically and graphically predictable that the convergence in time in our standing wave case would be of order higher than 4 after $t/T_{th}=100$ in our range of the $C_o$ parameter given the convergence rate of the amplitude and period at this time step.

It is then also consistent that the orders of convergence seem to increase with $\dfrac{t}{T_{th}}$ for all other computations.

\bibliographystyle{elsarticle-harv} 
\bibliography{ThesisBib_short}%

\end{document}